\documentclass[]{article}
\usepackage{amssymb}
\usepackage{amsmath,mathtools}
\usepackage{breqn}
\usepackage{cite}
\usepackage{stmaryrd}
\usepackage{latexsym}
\usepackage{graphicx}
\usepackage{subfigure}     
\usepackage{latexsym}
\usepackage{authblk}
\usepackage[misc]{ifsym}
\usepackage[papersize={20 cm, 24 cm},  text={17 cm,20 cm}, footskip=0.8 cm, headsep=0.8 cm, twoside,centering,showframe=false,showcrop=false]{geometry}
\usepackage{array}
\newcolumntype{P}[1]{>{\centering\arraybackslash}p{#1}}
\newcolumntype{M}[1]{>{\centering\arraybackslash}m{#1}}

\title{Bianchi-I cosmology with generalized Chaplygin gas and periodic deceleration parameter.}
\author[1]{N. Hulke}
\author[2]{G. P. Singh}
\author[3]{Binaya K. Bishi}

\affil[1]{\footnotesize Department of Applied Mathematics, Shri Ramdeobaba College of Engineering and Management, Nagpur, India.}
\affil[2]{\footnotesize Department of Mathematics, Visvesvaraya National Institute of Technology, Nagpur-440010, India.}
\affil[3]{\footnotesize Department of Mathematical Sciences, University of Zululand, Private Bag X1001, Kwa-Dlangezwa 3886, South Africa}
\affil[3]{\footnotesize Department of Mathematics, Lovely Professional University, Phagwara, Jalandhar, Punjab-144401, India, \footnote{binaybc@gmail.com}}

\begin{document}
	
	\maketitle
	
	\begin{abstract}
		We investigate the Bianchi-I cosmological model in presence of generalized Chaplygin gas (GCG), variable gravitational and cosmological constant. The exact solutions of Einstein field equations are obtained with time  varying periodic deceleration parameter. The graphical representation method has been used to discuss the physical and dynamical behaviour of the model. Further, the stability and physical acceptability of the obtained solutions have been investigated. Most of the parameters shows periodic behaviour in this study due to the presence of  cosine function in the deceleration parameter. In all the cases, pressure is negative, which leads us to late time expansion of the universe. The considered models are found to be stable.
	\end{abstract}

	\section{Introduction}
	Cosmological and astronomical data \cite{1,4,51,6,7,8,9} reveals that the universe is currently undergoing accelerating expansion and it has been originated with a bang from phase of very high density and temperature. For a long time, it was believed that either the universe will expand eternally or the inward pull of gravity, gradually slows down the expansion of the universe and would ultimately came to a halt and contract into a big crunch singularity. At the end of twentieth century, it has been discovered that the universe might be expanding with  acceleration. The fact of accelerating universe surprised the cosmologist as the idea of cosmic acceleration was against the standard predictions of decelerating expansion caused by gravity. The universe expansion is accelerating due to some exotic stuff termed as `Dark energy' (DE) having highly negative pressure. The successive disclosure in this direction gave more and more documentation for a flat, dark energy dominated accelerating universe \cite{14,2z,3z,4z,5z,6z,21,22z,n22,z23,z24,z25,z26,22,32a,32b,32c,32d,1z}. However, by modifying the theory of gravity one can find the alternative way to explain the acceleration of  expanding universe \cite{1z}.
	\par
	In General theory of relativity (GTR), cosmological constant $\Lambda$ had been introduced by Einstein, which now is one of the feasible candidates of dark energy \cite{10,11,12}. The homogeneous and isotropic model with cosmological constant in general relativistic framework is also known as Lambda cold dark matter ($\Lambda$CDM) model. However, the $\Lambda$CDM model faces some serious  issues \cite{13,14} such as the fine tuning (a typical small value), cosmic coincidence problems (although the universe is in accelerated phase of expansion, why the dark matter and the dark energy are of the same order?). To resolve these issues, various DE models with quintessence, k-essence, phantom, tachyon and so on \cite{1z}, were analysed in the literature. In spite of these attempts, the perfect DE model is still lacking \cite{1z}.
	\par Chaplygin gas (CG) \cite{15} with equation of state (EoS) $p=-\frac{B}{\rho}$ \cite{16} where, $\rho$ and $p$ are energy density and pressure respectively, is one of the most explored candidate of DE. Among the various class of dark energy models, CG is a simple characterisation in order to understand the cosmic acceleration and yields an unified scenario for dark energy and dark matter (DM) \cite{2z,3z,4z,5z,6z}. The remarkable feature of this unified model is that, at early times the CG behaves as a dust-like matter and at late times it behaves like cosmological constant. CG also plays a fascinating part in holography \cite{17} and string theory \cite{18} inspired models. CG depicts a transition to the present cosmic acceleration from a decelerated cosmic expansion and conceivably submit a deformation of $\Lambda$CDM model. CG model fails the tests connected with structure formation and observed strong oscillations of matter power spectrum \cite{19}. To overcome this failure of CG model, the generalised Chaplygin gas (GCG) with EoS $p=-\frac{B}{\rho^{\alpha}}$ with $0\leqslant \alpha \leqslant 1$ has been proposed \cite{2z,3z}. GCG model also acts like CG model i.e., at early times GCG mimic a dust-like matter and at late times it mimic a cosmological constant.
	With the aim of describing the unification of DE and DM, Zhang et al. \cite{4z} have proposed another version of GCG with EoS $p=-\frac{A(\mathcal{R})}{\rho^{\alpha}}$ ($\mathcal{R}$ is a scale factor) termed as new generalised Chaplygin gas (NGCG). By using various combinations of latest observational data sample including SNIa and CMB,  Salahedin et al. \cite{5z} have analysed for the constraints on free parameters of NGCG model by using the statistical Markov Chain Monte Carlo method. Mamon et al. \cite{6z} have  studied an extended GCG model with EoS of the NGCG and found that the NGCG model is consistent with gravitational thermodynamics. In literature, the GCG model is again modified as modified generalised Chaplygin gas (MGCG) model with EoS $p=\mu \rho-\frac{B}{\rho^\alpha}$, where $\mu$ is a positive constant \cite{4z}. This modification of GCG to MGCG has been done because the inferences from GCG models are almost similar to the CDM models \cite{21}. In general, above examples of cosmic fluid may written as $p=f(\rho)$ and we term these formulations as barotropic fluid. Variable Chaplygin gas (VCG) $p=-\frac{Ba^{-n}}{\rho}$ is one of the form among the class of modified forms of Chaplygin gas. A flat Friedmann-Lemaitre-Robertson-walker (FLRW) cosmological model with perfect fluid comprising of VCG yields $\Lambda$CDM model like behaviour at late-times \cite{n22}. In anisotropic Bianchi-I background, MCG equation of state and the barotropic fluid model satisfying general linear EoS may also characterize different phase of the universe and approach to $\Lambda$CDM model at late-times \cite{22z,z25}. Non-linear electrodynamics model may induce barotropic fluid like scenario  and results into varying deceleration parameter which is decreasing function of time \cite{z26}.
	\par
	In cosmological modelling, explanation of current accelerating stage and transition from decelerating past to accelerating stage of universe evolution are important and interesting ingredients. In order to explain the phase transition of the universe from decelerating to accelerating phase, one may also use varying deceleration parameter for anisotropic or isotropic universe model \cite{z23,z24}. Deceleration parameter denotes the rate at which the universe expansion is slowing down. The model of oscillating universe with quintom matter in FRW framework illustrate that the universe undergoes decelerating and accelerating expansion alternately and Hubble parameter oscillates and keeps positive for the time periodic varying deceleration parameter (TPVDP) \cite{22}. The field equations of Barber's second self-creation theory with TPVDP yield the values of state finder parameters $r$ and $s$ into state-finder parameters of standard $\Lambda$CDM model $r=1$ and $s=0$ for parameter $m=\frac{3}{2}$ \cite{23}. TPVDP model with string cloud in context of $f(R,T)$ gravity yield vanishing string tension density in the considered universe scenario \cite{24}. In the $f(R,T)$ gravity model, the cosmological solutions with TPVDP have been obtained for the flat homogenous and isotropic space-time and these solutions are consistent with observations for different values of model parameters  \cite{25}. Extension of FRW results in LRS  Bianchi-I space-time with $f(R,T)$ gravity yields the evolution of universe from decelerated phase to super exponential expansion \cite{27}. The late time acceleration of the universe may be caused in the presence of negative cosmological constant as a non conventional mechanism with TPVDP \cite{26}. Deceleration parameter has been used in modified gravity theories to investigate the cosmological implications along with other measurements in isotropic as well as anisotropic backgrounds \cite{26a,26e,26c,26b,26d}. The current accelerating expansion phase of universe may be explained in GTR framework with positive cosmological constant. The presence of cosmological constant in models may also affect the 	violation of energy conditions \cite{26f,26g}. In GTR framework, the Einstein field equations with the cosmological constant using variable redshift and shape functions have been solved for wormhole solutions \cite{26f}. The spherical regions for the wormhole may also satisfy the energy conditions for positive cosmological constant \cite{26f}. Energy conditions are coordinate invariant restrictions on	energy momentum tensor and force various linear combinations of energy density and pressure of model to be positive \cite{26f,26g,32e,29,30,31}.  The energy conditions have been used in literature, to derive many theorems such as the singularity theorems, black hole area increase theorem as well as the positive mass theorem \cite{29,30,31}.
	\par
	The introduction of anisotropies in the cosmological modelling may give rise to a richer dynamical structure, yet the model remains simple enough to provide numerical and/or analytical results. These anisotropic models allow us to analyse the problems regarding behaviour of models on the approach to space-time singularities like why do the present day universe appear highly isotropic, the effects of anisotropy on astronomical observables, etc \cite{27,z25,22z,1z,32e}. By considering the features of periodic deceleration parameter, role of Chaplygin gas and its modifications in cosmological modelling, it is worthwhile to investigate the dark energy cosmological model with time periodic varying deceleration parameter in Bianchi-I geometrical background. The paper is organized as follows: In section-2, we write the field equations of general relativity with Bianchi-I space-time metric. In section-3, we present the cosmological solutions by assuming the time periodic varying deceleration parameter and generalised Chaplygin gas. In section-4, we investigate the stability and acceptability of the solution using classical stability criterion, using the square sound speed and energy conditions. In section-5, we summarize our obtained result and concluding remarks.
	\section{Field equations}
	We take Bianchi-I space-time metric as,
	\begin{equation}
	\label{e1}
	ds^2=dt^2-\left(R_1^2(t)dx^2+R_2^2(t)dy^2+R_3^2(t)dz^2\right),
	\end{equation}
	where $R_1,\ R_2,\ R_3$ are directional scale factors. Einstein's field equations with gravitational and cosmological constant for perfect fluid distribution are given as
	\begin{equation}
	\label{e2}
	R_{ij}-\frac{1}{2}Rg_{ij}=-8\pi GT_{ij}+\Lambda g_{ij},
	\end{equation}
	where $R_{ij},\ g_{ij},\ R,\ G(t)$ and $\Lambda(t)$ are Ricci tensor, metric tensor, Ricci scalar, gravitational constant and cosmological constant respectively. $T_{ij}$ is energy momentum tensor and is given as
	\begin{equation}
	\label{e3}
	T_{ij}=(\rho+p)u_iu_j-pg_{ij},
	\end{equation}
	where $\rho$ is energy density, $p$ is perfect fluid pressure and $u^i$ represents the four velocity vector such that $u_iu^i=-1$. \\
	
	For the Bianchi-I space time metric represented by equation \eqref{e1}, the Einstein's field equation \eqref{e2} yields the following equations
	\begin{equation}
	\label{e4}
	\frac{\ddot{R_2}}{R_2}+\frac{\ddot{R_3}}{R_3}+\frac{\dot{R_2}}{R_2}\frac{\dot{R_3}}{R_3}=-8\pi Gp+\Lambda,
	\end{equation}
	\begin{equation}
	\label{e5}
	\frac{\ddot{R_1}}{R_1}+\frac{\ddot{R_3}}{R_3}+\frac{\dot{R_1}}{R_1}\frac{\dot{R_3}}{R_3}=-8\pi Gp+\Lambda,
	\end{equation}
	\begin{equation}
	\label{e6}
	\frac{\ddot{R_1}}{R_1}+\frac{\ddot{R_2}}{R_2}+\frac{\dot{R_1}}{R_1}\frac{\dot{R_2}}{R_2}=-8\pi Gp+\Lambda,
	\end{equation}
	\begin{equation}
	\label{e7}
	\frac{\dot{R_1}}{R_1}\frac{\dot{R_2}}{R_2}+\frac{\dot{R_2}}{R_2}\frac{\dot{R_3}}{R_3}+\frac{\dot{R_3}}{R_3}\frac{\dot{R_1}}{R_1}=8\pi G \rho+\Lambda,
	\end{equation}
	where overhead dot denotes derivative with respect to cosmic time $t$. From equations  \eqref{e4}-\eqref{e7} one can easily obtain
	\begin{equation}
	\label{e8}
	\dot{\rho}+(\rho+p)\left(\frac{\dot{R_1}}{R_1}+\frac{\dot{R_2}}{R_2}+\frac{\dot{R_3}}{R_3}\right)+\rho \frac{\dot{G}}{G}+\frac{\dot{\Lambda}}{8\pi G}=0.
	\end{equation}
	The energy momentum conservation equation $( T_{;j}^{ij}=0 )$ suggests                                                            \begin{equation}
	\label{e9}
	\dot{\rho}+(\rho+p)\left(\frac{\dot{R_1}}{R_1}+\frac{\dot{R_2}}{R_2}+\frac{\dot{R_3}}{R_3}\right)=0.
	\end{equation}
	From equations \eqref{e8} and \eqref{e9} we have
	\begin{equation}
	\label{e10}
	\rho \frac{\dot{G}}{G}+\frac{\dot{\Lambda}}{8\pi G}=0.
	\end{equation}
	Further, equations \eqref{e4}-\eqref{e6} yield the solutions
	\begin{equation}
	\label{e11}
	R_i=c_i\mathcal{R}e^{\frac{k_i}{3}\int\frac{1}{\mathcal{R}^3}dt},
	\end{equation}
	where $k_i$ and $c_i$ $(i = 1, 2, 3)$ are constants which satisfies $\sum_{i=1}^{3}k_i=0$ and $\prod_{i=1}^{3}c_i=1$. From the above results, one can notice that the metric potential can be explicitly expressed in terms of scale factor $\mathcal{R}=\left(R_1R_2R_3\right)^{\frac{1}{3}}$  represented by Bianchi-I space-time.
	\section{Cosmological Solutions}
	There are four linearly independent equations \eqref{e4}-\eqref{e7} with six unknowns in the form of $R_1$, $R_2$, $R_3$, $\rho$, $G$ and $\Lambda$. Hence to solve the system of equations completely, two additional physically plausible relations among these variables are required. To obtain the cosmological solution we considered the time periodically varying deceleration parameter (TPVDP) of the form \cite{22}
	\begin{equation}\label{e12}
	q=m\cos(nt)-1,
	\end{equation}
	where $m$ and $n$ are positive constants. This type of deceleration parameter is known as TPVDP. The deceleration parameter play a crucial role in determining the nature of the constructed models of the universe i.e.\ decelerating or accelerating in nature. According to the value/ranges of $q$ the universe exhibits the expansion in the following ways \cite{n22,31a}:
	\begin{itemize}
		\item $q>0$ : Decelerating expansion
		\item $q=0$ : Expansion with constant rate
		\item $-1<q<0$ : Accelerating power law expansion
		\item $q=-1$ : Exponential expansion/de Sitter expansion
		\item $q<-1$ : Super exponential expansion
	\end{itemize}
	From the considered form of $q$ in equation \eqref{e12}, the deceleration parameter shows periodic nature due to the presence of $\cos(nt)$. The deceleration parameter lies in the interval $-(m+1)\leq q \leq m-1$. Here we observed that:
	\begin{enumerate}
	\item For $m=0$, the deceleration parameter $q$ is equal to $-1$ and the universe exhibits exponential expansion/de Sitter expansion.
	\item For $m \in (0,1)$, the deceleration parameter $q$ becomes negative and leads to accelerated expansion in a periodic way.
	\item For $m=1$, $q$ lies in the interval $[-2,0]$, this shows that the universe evolves from expansion with constant rate to super exponential expansion in a periodic way followed by accelerating power law expansion to de Sitter expansion.
	\item For $m>1$, phase transition takes place from decelerating phase to accelerating phase in a periodic way where the universe starts with a decelerating expansion and evolves to super exponential expansion.
    \end{enumerate}
	The constraints on generlised deceleration parameter from cosmic chronometers were investigated and specifies the range as $q_0=-0.53^{+0.17}_{-0.13}$ \cite{30a}. In the study of particle creation mechnism in higher derivative theory, Singh et al. \cite{31a} examine the values of free parameters using the observatinal range of $q$. The present observational limit of the considered deceleration parameter $q_0=-0.53^{+0.17}_{-0.13}$ \cite{30a}, suggests the following values of $m$ and $n$, which is given in Table \ref{Tab11}.

	\begin{table}[ht!]
		\centering
		\caption{Computational range of the parameter $m$ for fixed values of $n$, obtained from the present value of the deceleration parameter.}
		{\begin{tabular}{@{}|c|c|@{}}
				\hline
				n & Interval of $m$ \\ \hline
				0.01 & $0.34 \leq m \leq 0.64$ \\ \hline
				0.02 & $0.35 \leq m \leq 0.66$ \\ \hline
				0.03 & $0.37 \leq m \leq 0.69$ \\ \hline
				0.04 & $0.39 \leq m \leq 0.74$ \\ \hline
				0.05 & $0.43 \leq m \leq 0.82$ \\ \hline
				0.06 & $0.49 \leq m \leq 0.94$ \\ \hline
				0.07 & $0.59 \leq m \leq 1.11$ \\ \hline
				0.08 & $0.74 \leq m \leq 1.39$ \\ \hline
				0.09 & $1.02 \leq m \leq 1.93$ \\ \hline
				0.10 & $1.70\leq m \leq 3.20$ \\ \hline
				\hline
			\end{tabular}}
			\label{Tab11}
		\end{table}
		
		\begin{figure}[ht!]
			\centering
			\begin{minipage}{.5\textwidth}
				\centering
				\includegraphics[width=.8\linewidth]{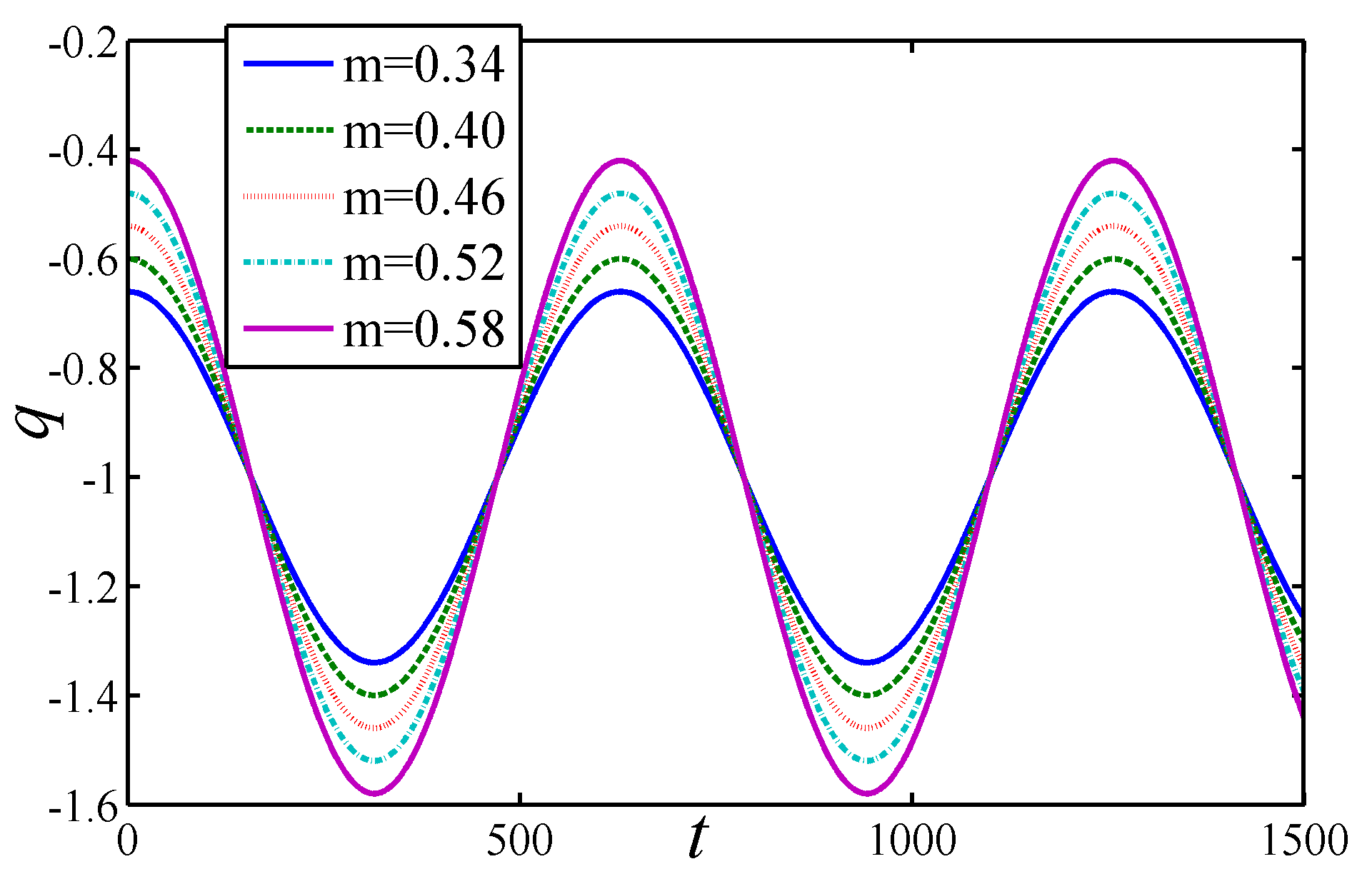}
				\caption{Deceleration parameter $q$ against \newline cosmic time for $0.34 \leq m \leq 0.64$ and $n=0.01$. }
				\label{fig1}
			\end{minipage}%
            \hfill
			\begin{minipage}{.5\textwidth}
				\centering
				\includegraphics[width=.8\linewidth]{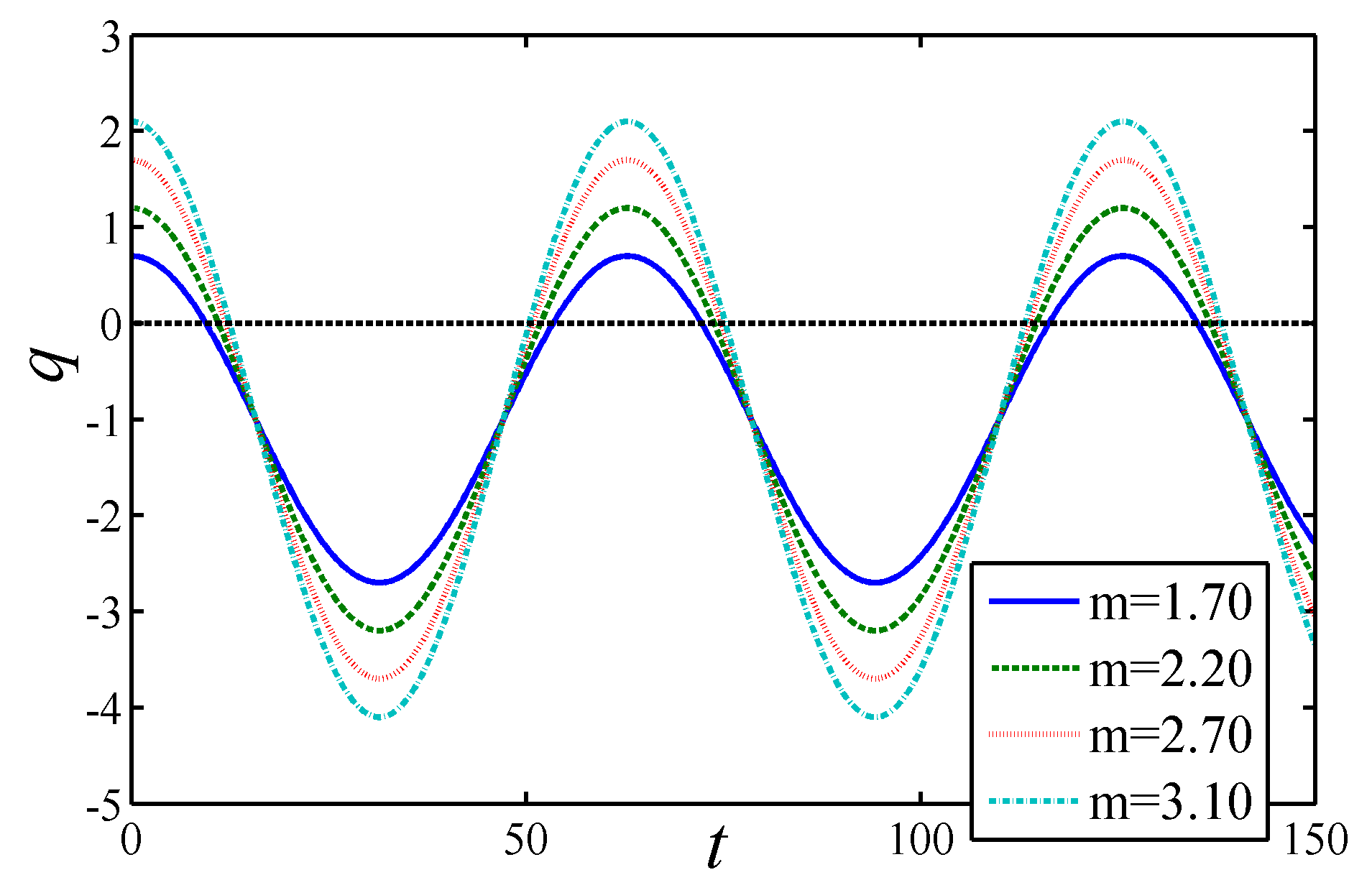}
				\caption{Deceleration parameter $q$ against \newline cosmic time for $1.70\leq m \leq 3.20$ and $n=0.10$.  }
				\label{fig2}
			\end{minipage}
		\end{figure}
		
		Figures (\ref{fig1}) and (\ref{fig2}) represents the behaviour of deceleration parameter for different values of $m$ and $n$. One can easily analyse the behaviour of $q$ from the plots. It can be seen that, (i) for $n=0.01$ and $0.34\leq m \leq 0.64$, considered models are accelerating in nature i.e $q<0$ and (ii) for $n=0.10$ and $1.70\leq m \leq 3.20$, models show phase transition from decelerating phase to accelerating phase. In this situation $q$ take values from positive to negative. We are mainly focus to investigate the phase transition scenario so in all discussed models, as a representative case we have considered the value of $m$ in the range  $1.70\leq m \leq 3.20$ and $n=0.10$.
		
		In order to obtain the Hubble parameter from equation \eqref{e12}, we used the relation between Hubble parameter and deceleration parameter as $q=\frac{d}{dt}(\frac{1}{H})-1$, which leads to
		\begin{equation}\label{e13}
		H=\frac{n}{m\sin(nt)+nc_4},\;c_4 \,\text{is the constant of integration}.
		\end{equation}
		The value of integration constant $c_4$ does not affect the qualitative behaviour of the Hubble parameter but it affect the scale factor. We classify the considered cosmological model in three different cases as per $c_4=0$, $c_4>0$ and $c_4<0,\:i.e.\ c_4=-c_5,\;c_5>0$.

		\subsection{Case-I: $c_4=0$}
		In this case the Hubble parameter in equation \eqref{e13} leads to
		\begin{equation}\label{e14}
		H=\frac{n}{m\sin(nt)}.
		\end{equation}
		We know the relation between Hubble parameter and scale factor as $H=\frac{\dot{\mathcal{R}}}{\mathcal{R}}$, which along with equation \eqref{e14} leads to the scale factor of the form
		\begin{equation}\label{e15}
		\mathcal{R}=c_6\left[\frac{1-\cos(nt)}{\sin(nt)}\right]^{\frac{1}{m}}=c_6\left[\tan\left(\frac{nt}{2}\right)\right]^{\frac{1}{m}},\:\:c_6\; \text{is the constant of integration}.
		\end{equation}
		Further, equations \eqref{e11} and \eqref{e15} yields the following metric potentials as
		\begin{equation}\label{e16}
		R_i=c_{i1}\left[\tan\left(\frac{nt}{2}\right)\right]^{\frac{1}{m}}exp\left[{c_{i2}\int \left[\tan\left(\frac{nt}{2}\right)\right]^{-\frac{3}{m}}dt}\right],
		\end{equation}
		where $c_{i1}=c_ic_6$ and $c_{i2}=\frac{k_i}{3c_6^3}$ for $i=1,2,3$.
		We take the generalized Chaplygin gas, described by EoS \cite{2z,3z}
		\begin{equation}\label{e17}
		p=-\frac{A}{\rho^\alpha},\; A>0\:\text{and}\: 0\leq\alpha \leq 1.
		\end{equation}
		The expression for the energy density can be obtained from equations \eqref{e9},\eqref{e14} and \eqref{e17} as
		\begin{equation}\label{e18}
		\rho=\left[A+\frac{c_7}{\mathcal{R}^{3(1+\alpha)}}\right]^{\frac{1}{1+\alpha}}=\left[A+\rho_0\left(\sin(nt)\right)^{-\frac{3(1+\alpha)}{m}}\left(1+\cos(nt)\right)^{\frac{3(1+\alpha)}{m}}\right]^{\frac{1}{1+\alpha}},
		\end{equation}
		where $\rho_0=c_7c_6^{-3(1+\alpha)}$ and $c_7$ is the constant of integration.
		The directional Hubble parameter $H_i$ are given by,
		\begin{equation}\label{e19}
		H_i=\frac{\dot{R}_i}{{R}_i}=\frac{3c_6^3n+\left(\sin(nt)\right)^{\frac{m-3}{m}}\left(1+\cos(nt)\right)^{\frac{3}{n}}k_im}{3c_6^3m\sin(nt)},\;i=1,2,3.
		\end{equation}
		From Equations \eqref{e4},\eqref{e7} , \eqref{e16}, \eqref{e17} and \eqref{e18}, one can get the expression for Gravitational constant as
		\begin{equation}\label{e20}
		G=\frac{1}{8\pi(\rho+p)}\left[\frac{k_4\left(\sin(nt)\right)^{-\frac{6}{m}}(\cos(nt)-1)\left(\cos(nt)+1\right)^{\frac{m+6}{m}}-18c_6^6n^2\cos(nt)}{9c_6^6m(\cos^2(nt)-1)}\right],
		\end{equation}
		where $k_4=m(k_1k_2+k_1k_3-k_2^2-k_3^2)$.
		With the help of equations \eqref{e4} and \eqref{e7}, one can obtain the expression for cosmological constant as
		\begin{equation}\label{e21}
		\Lambda=\frac{1}{\rho+p}\bigg[\frac{\splitfrac{m^2\left(\sin(nt)\right)^{-\frac{6}{m}}(\cos(nt)-1)\left(\cos(nt)+1\right)^{\frac{m+6}{m}}(k_5\rho+k_6p)}{-27c_6^6n^2(\rho+p)+18c_6^6n^2m\cos(nt)\rho}}{9c_6^6m^2(\cos^2(nt)-1)}\bigg],
		\end{equation}
		where $k_5=k_2k_3+k_2^2+k_3^2$ and $k_6=k_1k_2+k_1k3+k_2k_3$.\\
		The physical quantities of the observational interest are expansion scalar ($\Theta$), shear scalar ($\sigma^2$) and the anisotropic parameter ($A_m$), which are defined as follows:
		\begin{equation}\label{e22}
		\Theta=3H=H_1+H_2+H_3,
		\end{equation}
		\begin{equation}\label{e23}
		\sigma^2=\frac{1}{2}\left[H_1^2+H_2^2+H_3^2\right]-\frac{\Theta^2}{6},
		\end{equation}
		\begin{equation}\label{e24}
		A_m=\frac{1}{3}\sum_{i=1}^{3}\left(\frac{H_i-H}{H}\right)^2.
		\end{equation}
		In this case, the physical quantities are obtained as
		\begin{equation}\label{e25}
		\Theta=\frac{3n}{m\sin(nt)},
		\end{equation}
		\begin{equation}\label{e26}
		\sigma^2=\frac{(k_1^2+k_2^2+k_3^2)\left[1+\cos(nt)\right]^{\frac{6}{m}}\left[\sin(nt)\right]^{\frac{-6+2m}{m}}}{18c_6^6(1-\cos^2(nt))}=\frac{k_1^2+k_2^2+k_3^2}{18c_6^6}\left[\frac{1+\cos(nt)}{\sin(nt)}\right]^{\frac{6}{m}},
		\end{equation}
		\begin{equation}\label{e27}
		A_m=\frac{(k_1^2+k_2^2+k_3^2)m^2\left[1+\cos(nt)\right]^{\frac{6}{m}}\left[\sin(nt)\right]^{\frac{-6+2m}{m}}}{27c_6^6n^2}.
		\end{equation}
		In this case the state finder parameters are defined and expressed as
		\begin{equation}\label{e28}
		r=\frac{\dddot{\mathcal{R}}}{\mathcal{R}H^3}=m^2\cos^2(nt)-3m\cos(nt)+1+m^2,
		\end{equation}
		\begin{equation}\label{e29}
		s=\frac{r-1}{3(q-0.5)}=\frac{2m(m\cos^2(nt)-3\cos(nt)+m)}{3(2m\cos(nt)-3)}.
		\end{equation}
		Relation between $r$ and $s$ is given by
		\begin{equation}\label{e30}
		r=1+\frac{9}{2}s^2\pm \frac{3}{2}s\sqrt{9+9s^2-4m^2}.
		\end{equation}
		In terms of the deceleration parameter the state finder parameters are given as
		\begin{equation}\label{e31}
		r=-1-q+q^2+m^2,
		\end{equation}
		\begin{equation}\label{e32}
		s=\frac{2(-2-q+q^2+m^2)}{3(-1+2q)}.
		\end{equation}
		
		\begin{table}[ht!]
			\centering
			\caption{\{r,s\}=(1,0) corresponding to the different values of $m$ and $q$.}
			{\begin{tabular}{@{}|c|c|c|@{}}
					\hline
					$q$ & $m$ & \{r,s\} pair \\ \hline
					-1.00& 	 0.000& (1.000,0)\\ \hline
					-0.80& 	 0.748& (1.0,2.846726e-017)$\approx $(1,0)\\\hline
					-0.60& 	 1.020& (1.0,0)\\ \hline
					-0.40& 	 1.200& (1.0,-8.223874e-017)$\approx $ (1,0)\\ \hline
					-0.20& 	 1.327& (1.0,1.057355e-016)$\approx $ (1,0)\\ \hline
					0.00& 	 1.414& (1.0,-2.960595e-016)$\approx $ (1,0)\\ \hline
					0.20& 	 1.470& (1.0,0)\\ \hline
					0.40& 	 1.497& (1.0,0) \\ \hline
					0.60& 	 1.497& (1.0,0)\\ \hline
					0.80&	 1.470&	 (1.0,0)\\ \hline
					1.00&	 1.414&	 (1.0,2.960595e-016)$\approx $ (1,0)\\
					\hline
				\end{tabular}}
				\label{Tab12}
			\end{table}
			The energy density, pressure, cosmological constant and gravitational constant are periodic in nature, which is noticed from the Figure (\ref{fig4}) to Figure (\ref{fig7}) respectively due to the presence of $\cos(nt)$ and $\sin(nt)$ terms in the expressions of these physical quantities. Here one can note that, $\rho, \;p,\; \Lambda,\; G\rightarrow\infty$ at $t=\frac{n_2\pi}{n},\;\forall \;n_2\in \textbf{Z}$. The energy density is positive where as pressure is negative for different values of $m$ with evolvement of time. In this case, the qualitative behavior of energy density follow the pattern of higher energy density value to lower energy density value (approaching to zero) to higher energy density value. This process will continue due to the periodic nature of the terms $\cos(nt)$ and $\sin(nt)$ involve in the expression of energy density \eqref{e18}. Again it is also pointed out that, cosmological constant is positive for some values of $m$ and positive  to negative values for some $m$ (See Figure \ref{fig6}). Gravitational constant takes values from negative to positive values and positive to negative values in a periodic way for different values of $m$ (See Figure \ref{fig7}). Further, it is noticed that the expansion scalar, shear scalar and anisotropy parameter also have singularity at $t=\frac{n_2\pi}{n},\;\forall \;n_2\in \textbf{Z}$ and they behaves periodically.
			\begin{figure}[ht!]
				\centering
				\includegraphics[width=.8\linewidth]{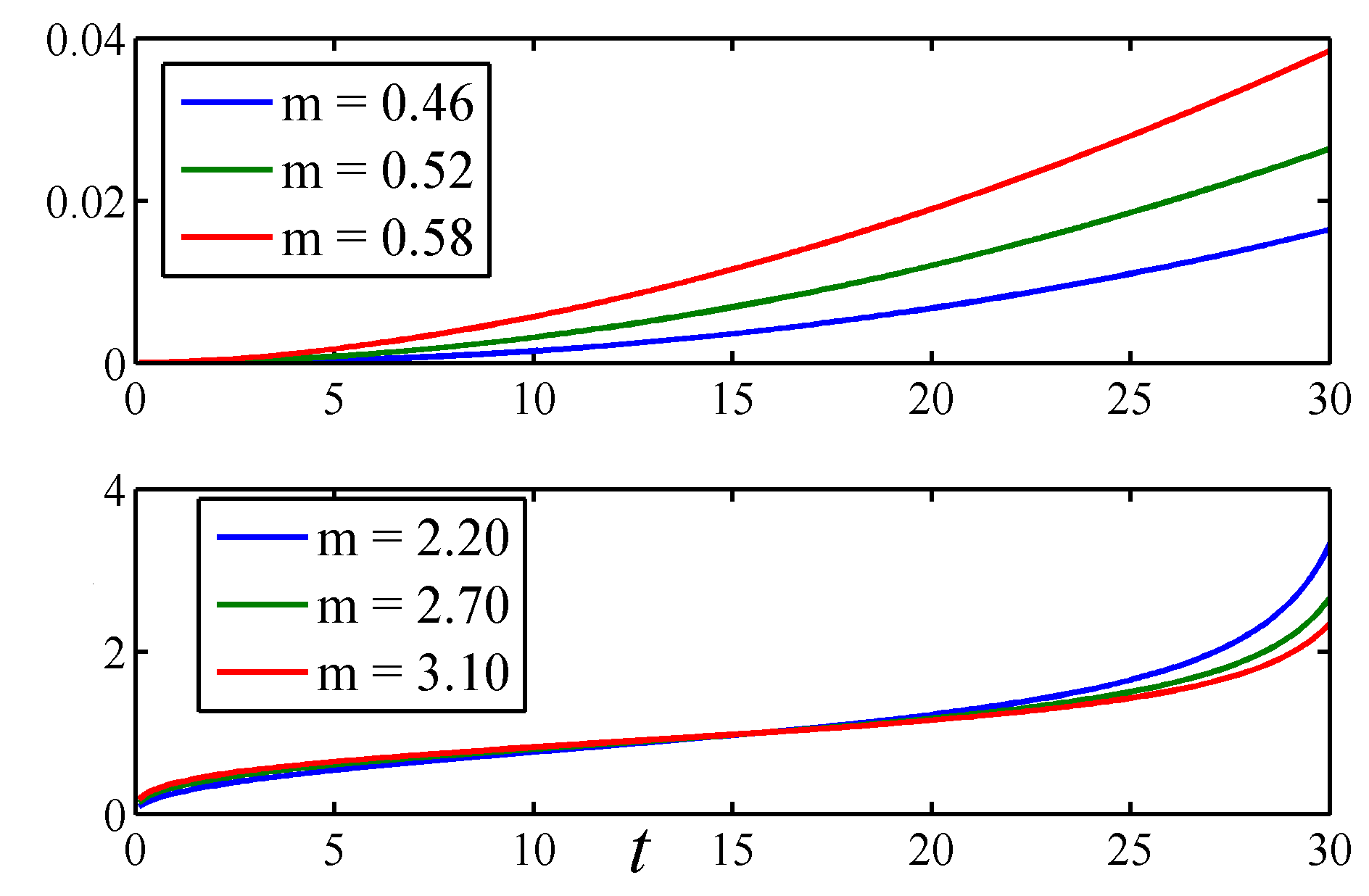}
				\caption{Scale factor parameter $\mathcal{R}$ against cosmic time for different $m$ and $n$.}
				\label{fig3}
			\end{figure}
			
			\begin{figure}[ht]
				\centering
				\begin{minipage}{.5\textwidth}
					\centering
					\includegraphics[width=.8\linewidth]{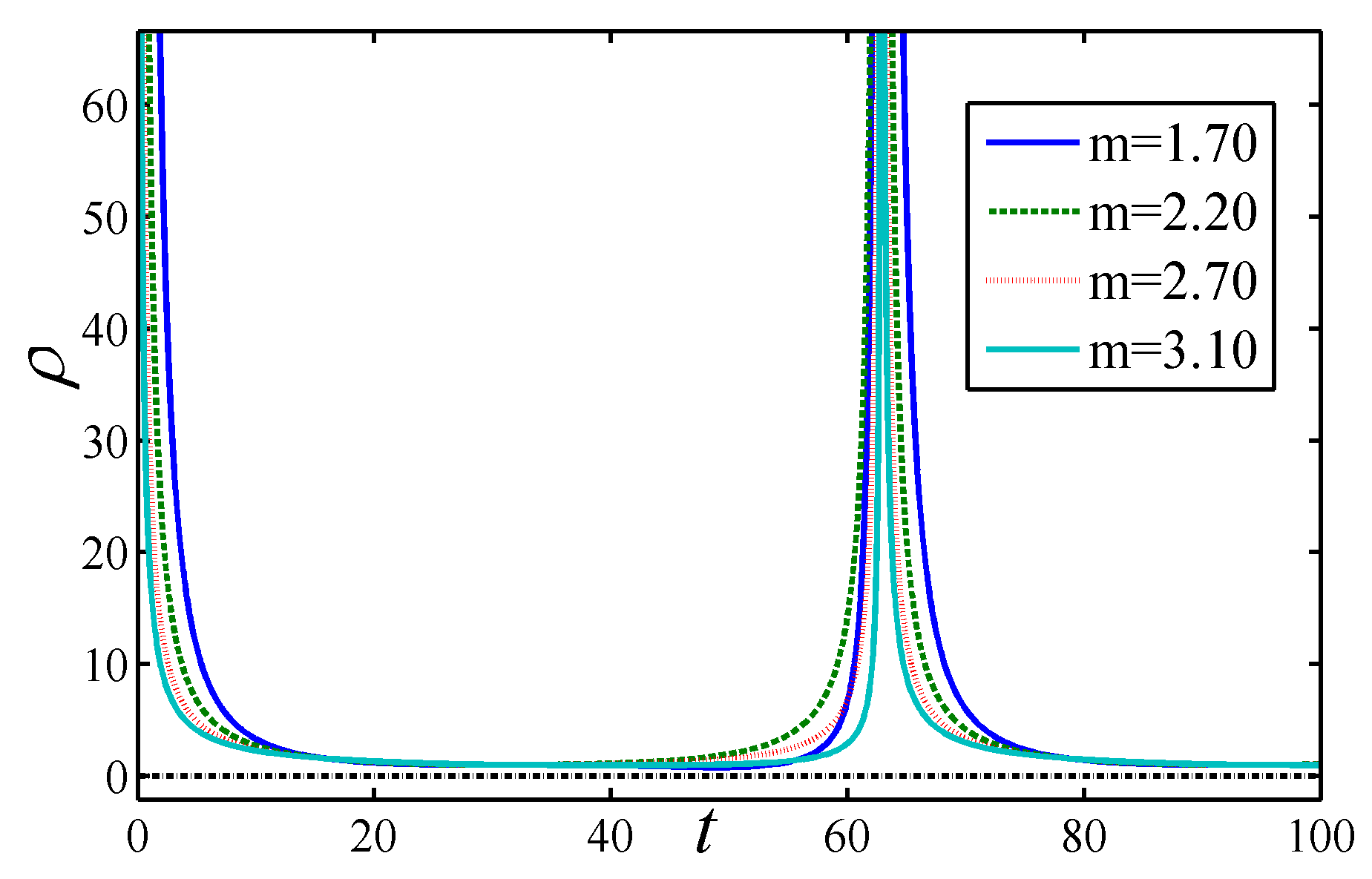}
					\caption{Energy density $\rho$ against cosmic time  \newline for different $m$, $n=0.1$, $\alpha=0.5$, $\rho_0=1$ and $A=1$.}
					\label{fig4}
				\end{minipage}%
				\begin{minipage}{.5\textwidth}
					\centering
					\includegraphics[width=.8\linewidth]{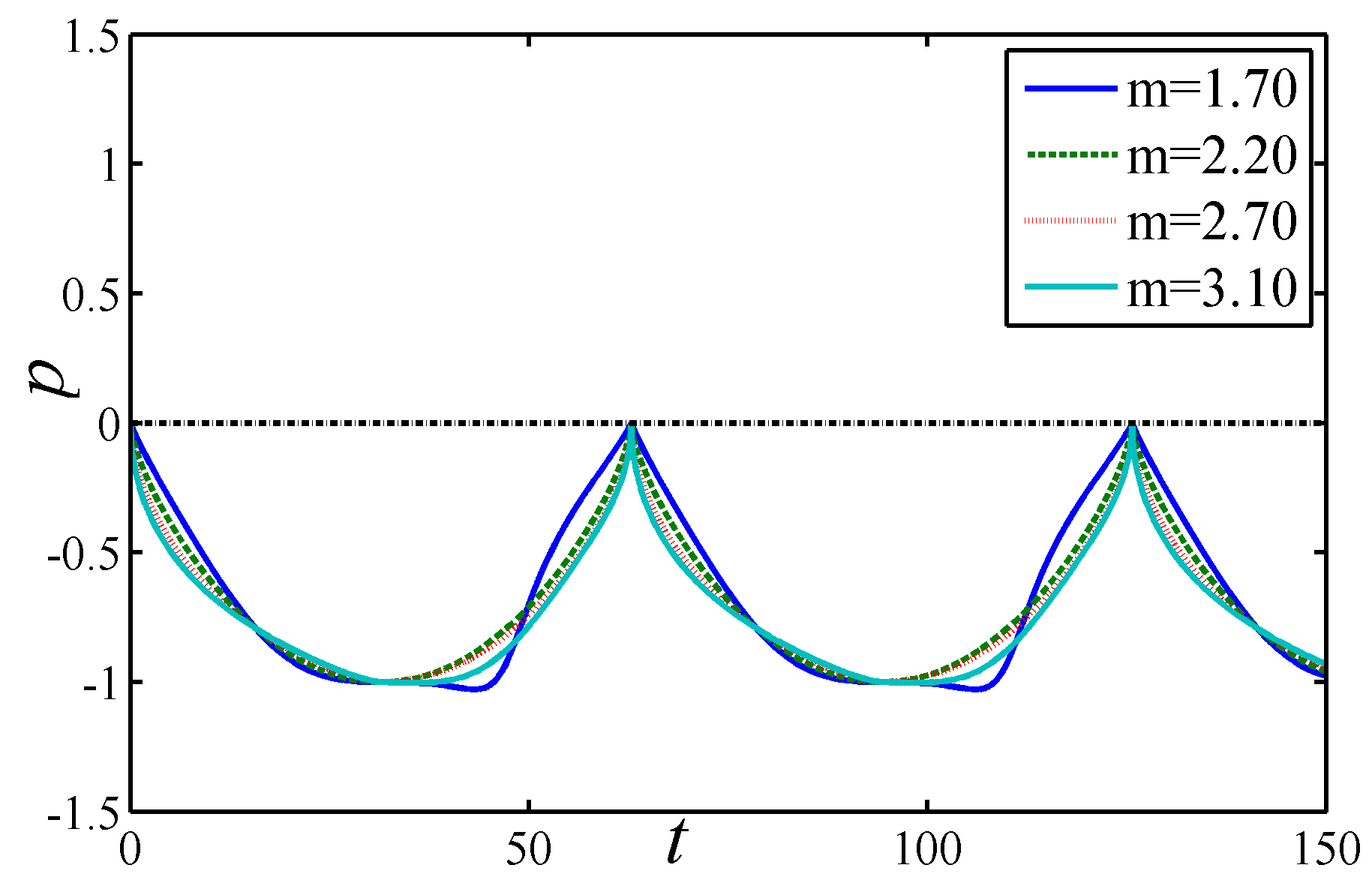}
					\caption{Pressure $p$ against cosmic time for different $m$, $n=0.1$, $\alpha=0.5$, $\rho_0=1$ and $A=1$.}
					\label{fig5}
				\end{minipage}
			\end{figure}
			The profile of scale factor $\mathcal{R}$, energy density $\rho$, pressure $p$, cosmological constant $\Lambda$ and gravitational constant $G$ against time is presented in the Figure (\ref{fig3}) to Figure (\ref{fig7}) respectively for fix $n$ and different values of $m$ with suitable choice of arbitrary constant involve in the expressions of the physical quantities. The scale factor is increasing with the evolvement of time in the provided range of time, which can be seen from the Figure (\ref{fig3}). However, the qualitative behaviour is similar to $\tan$ function due to the presence of $\tan\left(\frac{nt}{2}\right)$ in equation \eqref{e15} and $\mathcal{R}\rightarrow \infty$ at $t=\frac{(2n_1+1)\pi}{n},\; n_1 \in \textbf{Z}$.
			\begin{figure}[ht!]
				\centering
				\begin{minipage}{.5\textwidth}
					\centering
					\includegraphics[width=.8\linewidth]{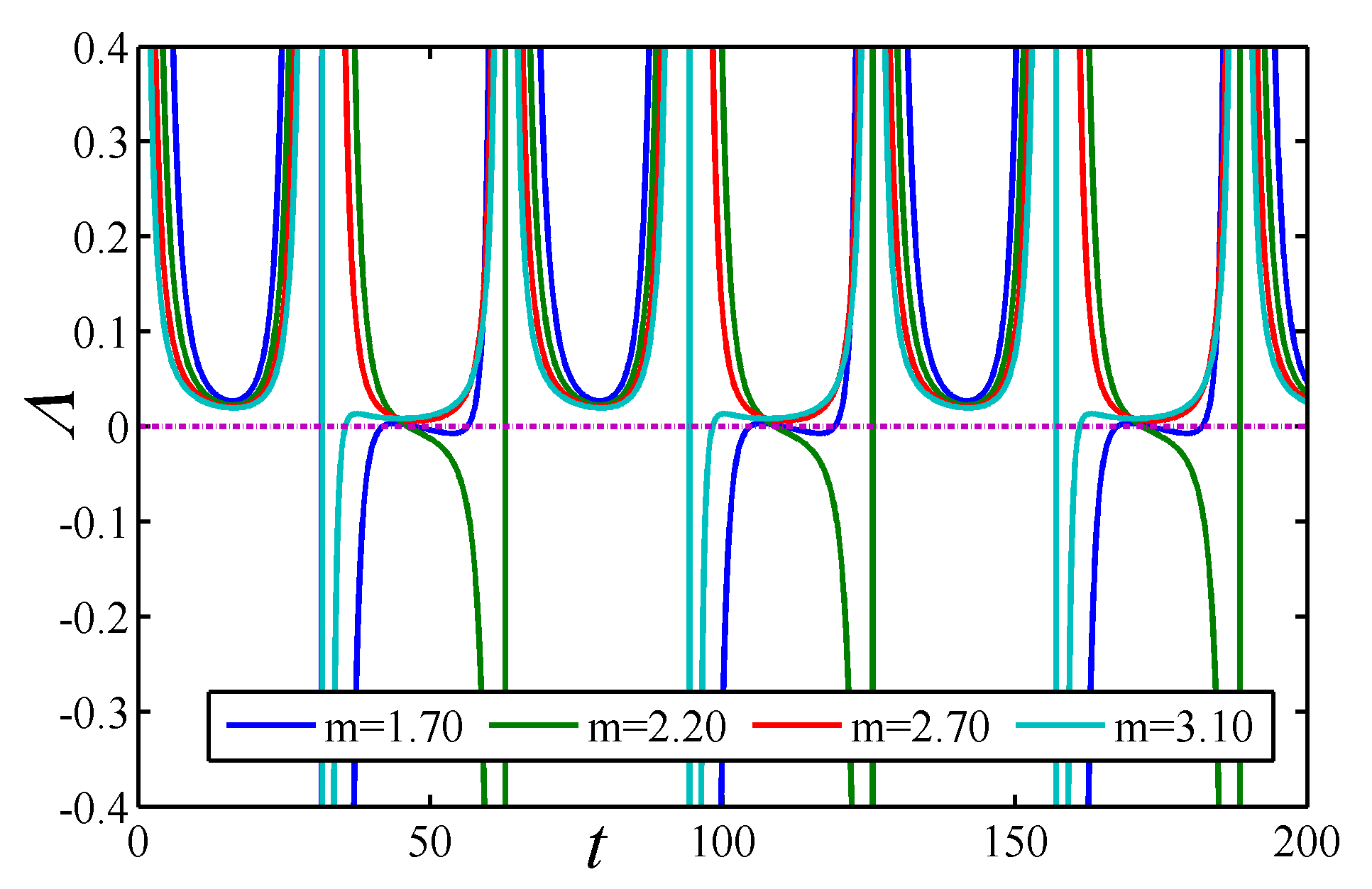}
					\caption{Cosmological constant $\Lambda$ against cosmic \newline time for different $m$, $n=0.1$, $\alpha=0.5$, and $A=1$.}
					\label{fig6}
				\end{minipage}%
				\begin{minipage}{.5\textwidth}
					\centering
					\includegraphics[width=.8\linewidth]{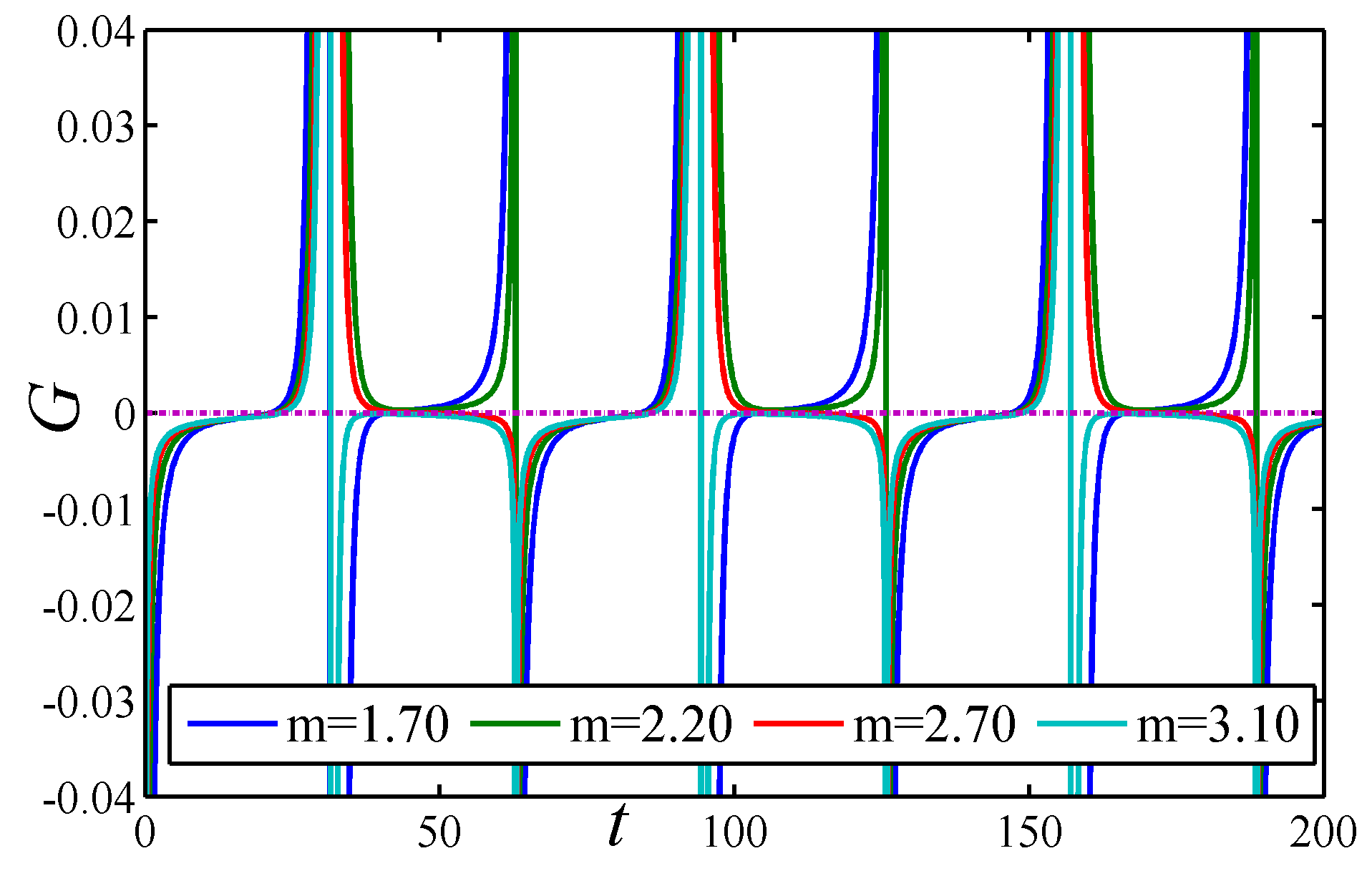}
					\caption{Gravitational constant $G$ against cosmic time for different $m$, $n=0.1$, $\alpha=0.5$  and $A=1$.}
					\label{fig7}
				\end{minipage}
			\end{figure}

			\subsection{Case-II: $c_4>0$}
\begin{figure}[ht!]
				\centering
				\includegraphics[width=.8\linewidth]{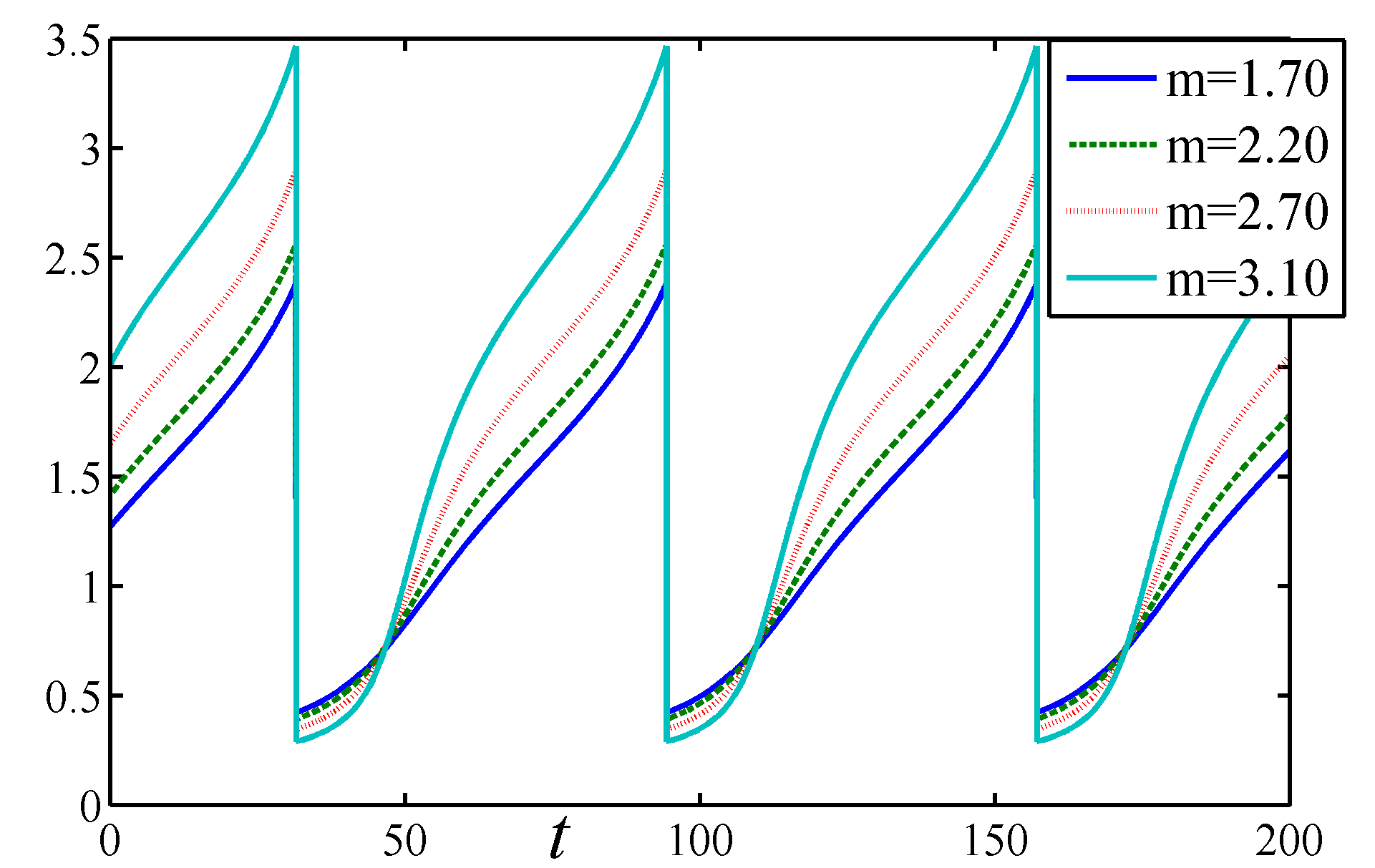}
				\caption{Scale factor $\mathcal{R}$ against cosmic time for different $m$ }
				\label{fig8}
			\end{figure}

			In this case the Hubble parameter in equation \eqref{e13} takes the form
			\begin{equation}\label{e33}
			H=\frac{n}{m\sin(nt)+nc_4}.
			\end{equation}
			The relation $H=\frac{\dot{\mathcal{R}}}{\mathcal{R}}$, which along with \eqref{e33} leads to the scale factor of the form
			\begin{equation}\label{e34}
			\mathcal{R}=c_8e^{\frac{2}{\sqrt{n^2c_4^2-m^2}}\arctan\left(\frac{nc_4\tan\left(\frac{nt}{2}\right)+m}{\sqrt{n^2c_4^2-m^2}}\right)},\:\:c_8\; \text{is the constant of integration}.
			\end{equation}
			Further, equations \eqref{e11} and \eqref{e34} yields the following metric potentials as
			\begin{equation}\label{e35}
		{R}_i=c_{i3} e^{\frac{2\arctan\left(\frac{nc_4\tan\left(\frac{nt}{2}\right)+m}{\sqrt{n^2c_4^2-m^2}}\right)}{\sqrt{n^2c_4^2-m^2}}}
			exp\left[c_{i4}\int e^{-\frac{6\arctan\left(\frac{nc_4\tan\left(\frac{nt}{2}\right)+m}{\sqrt{n^2c_4^2-m^2}}\right)}{\sqrt{n^2c_4^2-m^2}}}dt\right],\; i=1,2,3,
			\end{equation}
			where $c_{i3}=c_ic_8$ and $c_{i4}=\frac{k_i}{3c_8^3} \: (i=1,2,3)$. The directional Hubble parameters are expressed as
			\begin{equation}\label{e36}
			H_i=\frac{\left[3nc_8^3e^{\frac{6\arctan\left(\frac{nc_4\tan\left(\frac{nt}{2}\right)+m}{\sqrt{n^2c_4^2-m^2}}\right)}{\sqrt{n^2c_4^2-m^2}}}+nc_4k_i+mk_i\sin(nt)\right]e^{-\frac{6\arctan\left(\frac{nc_4\tan\left(\frac{nt}{2}\right)+m}{\sqrt{n^2c_4^2-m^2}}\right)}{\sqrt{n^2c_4^2-m^2}}}}{3c_8^3(nc_4+m\sin(nt))}.
			\end{equation}
			The expression for the energy density can be obtained from equations \eqref{e9} and \eqref{e34} as
			\begin{equation}\label{e37}
			\rho=\left[A+\frac{c_7}{\mathcal{R}^{3(1+\alpha)}}\right]^{\frac{1}{1+\alpha}}=\left[A+\rho_1e^{-\frac{6(1+\alpha)\arctan\left(\frac{nc_4\tan\left(\frac{nt}{2}\right)+m}{\sqrt{n^2c_4^2-m^2}}\right)}{\sqrt{n^2c_4^2-m^2}}}\right]^{\frac{1}{1+\alpha}},
			\end{equation}
			where $\rho_1=c_7c_8^{-3(1+\alpha)}$. The expression for gravitational and cosmological constant are obtained as
			\begin{equation}\label{e38}
			G=\frac{\splitfrac{-9n^2mc_8^6\left(\cos^2\left(\frac{nt}{2}\right)-\frac{1}{2}\right)e^{\frac{12\arctan\left(\frac{nc_4\tan\left(\frac{nt}{2}\right)+m}{\sqrt{n^2c_4^2-m^2}}\right)}{\sqrt{n^2c_4^2-m^2}}}+\bigg(m^2\cos^4\left(\frac{nt}{2}\right)}{-m^2\cos^2\left(\frac{nt}{2}\right)-nmc_4\sin\left(\frac{nt}{2}\right)\cos\left(\frac{nt}{2}\right)-\frac{n^2c_4^2}{4}}\bigg)k_7}{72\pi c_8^6(\rho+p)\left(\splitfrac{m^2\cos^4\left(\frac{nt}{2}\right)-m^2\cos^2\left(\frac{nt}{2}\right)}{-nmc_4\sin\left(\frac{nt}{2}\right)\cos\left(\frac{nt}{2}\right)-\frac{n^2c_4^2}{4}}\right)e^{\frac{12\arctan\left(\frac{nc_4\tan\left(\frac{nt}{2}\right)+m}{\sqrt{n^2c_4^2-m^2}}\right)}{\sqrt{n^2c_4^2-m^2}}}},
			\end{equation}
			where $k_7=\frac{k_4}{m}$,
			\begin{equation}\label{e39}
			\Lambda=\frac{\splitfrac{9n^2c_8^6\left(\rho m\left(\cos^2\left(\frac{nt}{2}\right)-\frac{1}{2}\right)-\frac{3}{4}(\rho+p)\right)e^{\frac{12\arctan\left(\frac{nc_4\tan\left(\frac{nt}{2}\right)+m}{\sqrt{n^2c_4^2-m^2}}\right)}{\sqrt{n^2c_4^2-m^2}}}}{+\bigg(m^2\cos^4\left(\frac{nt}{2}\right)-m^2\cos^2\left(\frac{nt}{2}\right)-nmc_4\sin\left(\frac{nt}{2}\right)\cos\left(\frac{nt}{2}\right)-\frac{n^2c_4^2}{4}\bigg)(k_5\rho+k_6 p)}}{9c_8^6(\rho+p)\left(\splitfrac{m^2\cos^4\left(\frac{nt}{2}\right)-m^2\cos^2\left(\frac{nt}{2}\right)}{-nmc_4\sin\left(\frac{nt}{2}\right)\cos\left(\frac{nt}{2}\right)-\frac{n^2c_4^2}{4}}\right)e^{\frac{12\arctan\left(\frac{nc_4\tan\left(\frac{nt}{2}\right)+m}{\sqrt{n^2c_4^2-m^2}}\right)}{\sqrt{n^2c_4^2-m^2}}}}.
			\end{equation}
			In this case, the physical quantities are obtained as
			\begin{equation}\label{e40}
			\Theta=\frac{3n}{m\sin(nt)+nc_4},
			\end{equation}
			\begin{equation}\label{e41}
			\sigma^2=\frac{(k_1^2+k_2^2+k_3^2)}{18c_8^6}e^{-\frac{12\arctan\left(\frac{nc_4\tan\left(\frac{nt}{2}\right)+m}{\sqrt{n^2c_4^2-m^2}}\right)}{\sqrt{n^2c_4^2-m^2}}},
			\end{equation}
			\begin{equation}\label{e42}
			A_m=-\frac{(k_1^2+k_2^2+k_3^2)(m^2\cos^2(nt)-2nmc_4\sin(nt)-m^2-n^2c_4^2)}{27n^2c_8^6}e^{-\frac{12\arctan\left(\frac{nc_4\tan\left(\frac{nt}{2}\right)+m}{\sqrt{n^2c_4^2-m^2}}\right)}{\sqrt{n^2c_4^2-m^2}}}.
			\end{equation}
			In this case the state finder parameters are defined and expressed as
			\begin{equation}\label{e43}
			r=\frac{\dddot{\mathcal{R}}}{\mathcal{R}H^3}=-\frac{(m\sin(nt)+nc_4)^3\bigg(\splitfrac{1+3m+2m^2+nmc_4\sin(nt)}{+4m^2\cos^4\left(\frac{nt}{2}\right)-(4m^2+6m)\cos^2\left(\frac{nt}{2}\right)\bigg)}}{\splitfrac{\left(12nm^2c_4+4m^3\sin(nt)\right)\cos^4\left(\frac{nt}{2}\right)-\left(12nm^2c_4+4m^3\sin(nt)\right)\cos^2\left(\frac{nt}{2}\right)}{-3n^2mc_4^2\sin(nt)-n^3c_4^3}},
			\end{equation}
			\begin{eqnarray}\label{e44}
				s&=&\frac{r-1}{3(q-0.5)}=-\frac{2}{3(2m\cos(nt)-3)}\nonumber\\
				&\times&\left[\frac{(m\sin(nt)+nc_4)^3\bigg(\splitfrac{1+3m+2m^2+nmc_4\sin(nt)}{+4m^2\cos^4\left(\frac{nt}{2}\right)-(4m^2+6m)\cos^2\left(\frac{nt}{2}\right)\bigg)}}{\splitfrac{\left(12nm^2c_4+4m^3\sin(nt)\right)\cos^4\left(\frac{nt}{2}\right)-\left(12nm^2c_4+4m^3\sin(nt)\right)\cos^2\left(\frac{nt}{2}\right)}{-3n^2mc_4^2\sin(nt)-n^3c_4^3}}+1\right].
			\end{eqnarray}
			In terms of the deceleration parameter the state finder parameters are given as
			\begin{equation}\label{e45}
			r=\frac{\left(\sqrt{m^2-1-2q-q^2}+nc_4\right)^3\left(m^2-1-q+q^2+nc_4\sqrt{m^2-1-2q-q^2}\right)}{\left(m^2-1-2q-q^2+3n^2c_4^2\right)\sqrt{m^2-1-2q-q^2}+3nc_4\left(m^2-1-2q-q^2+\frac{n^2c_4^2}{3}\right)},
			\end{equation}
			\begin{eqnarray}\label{e46}
				s&=&\frac{2\left(\sqrt{m^2-1-2q-q^2}+nc_4\right)^3\left(m^2-1-q+q^2+nc_4\sqrt{m^2-1-2q-q^2}\right)}{3(2q-1)\left(\left(m^2-1-2q-q^2+3n^2c_4^2\right)\sqrt{m^2-1-2q-q^2}+3nc_4\left(m^2-1-2q-q^2+\frac{n^2c_4^2}{3}\right)\right)}\nonumber\\
				&-&\frac{2}{3(2q-1)}.
			\end{eqnarray}
			Figure (\ref{fig8}) indicates the profile of scale factor against the cosmic time. Here we noticed that, the scale factor is increasing with respect to cosmic time in a periodic way.
			\par
\begin{figure}[ht!]
				\centering
				\begin{minipage}{.5\textwidth}
					\centering
					\includegraphics[width=.8\linewidth]{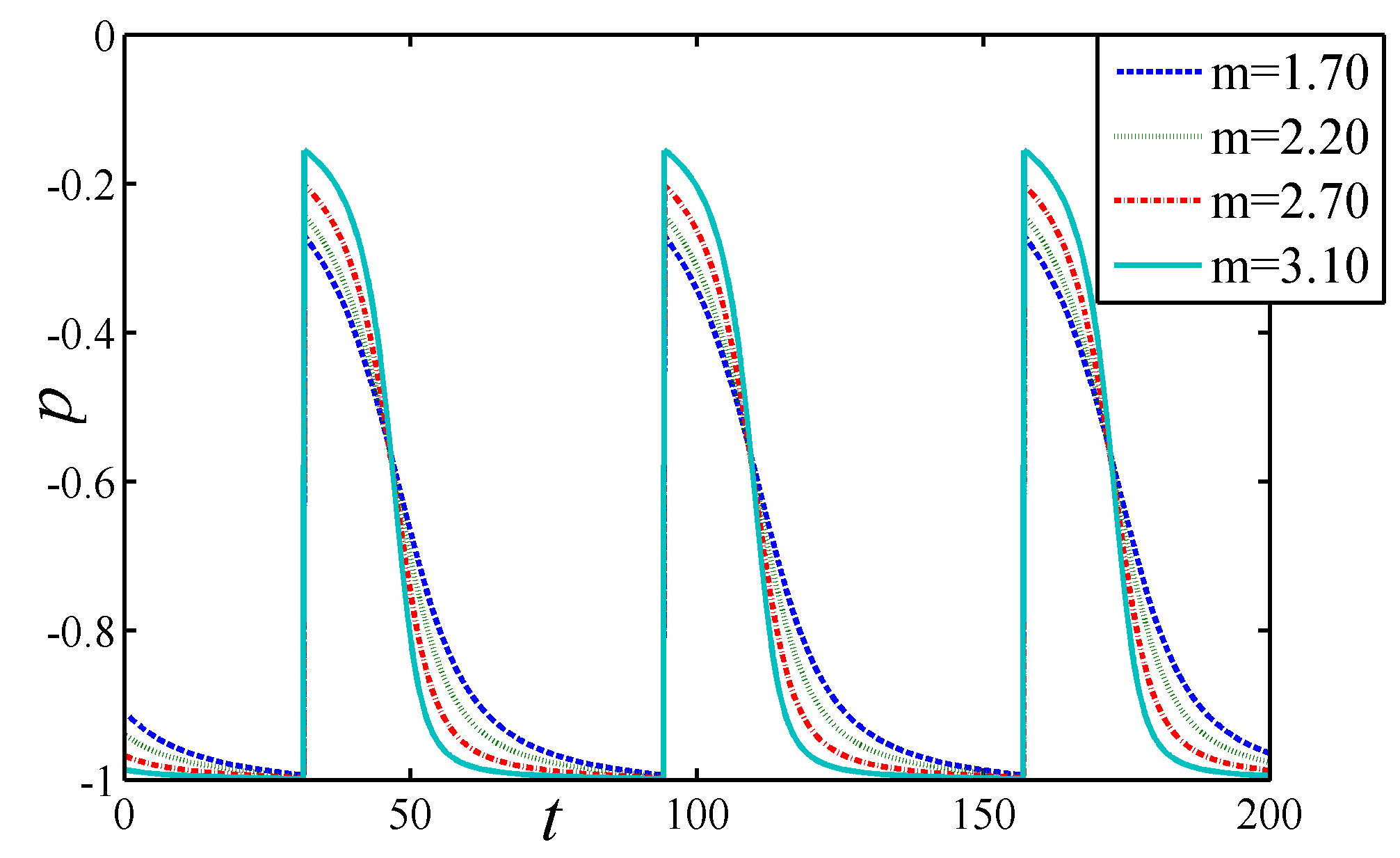}
					\caption{Pressure $p$ against cosmic time for \newline $1.70\leq m \leq 3.20$ and $n=0.10$. }
					\label{fig9}
				\end{minipage}%
				\begin{minipage}{.5\textwidth}
					\centering
					\includegraphics[width=.8\linewidth]{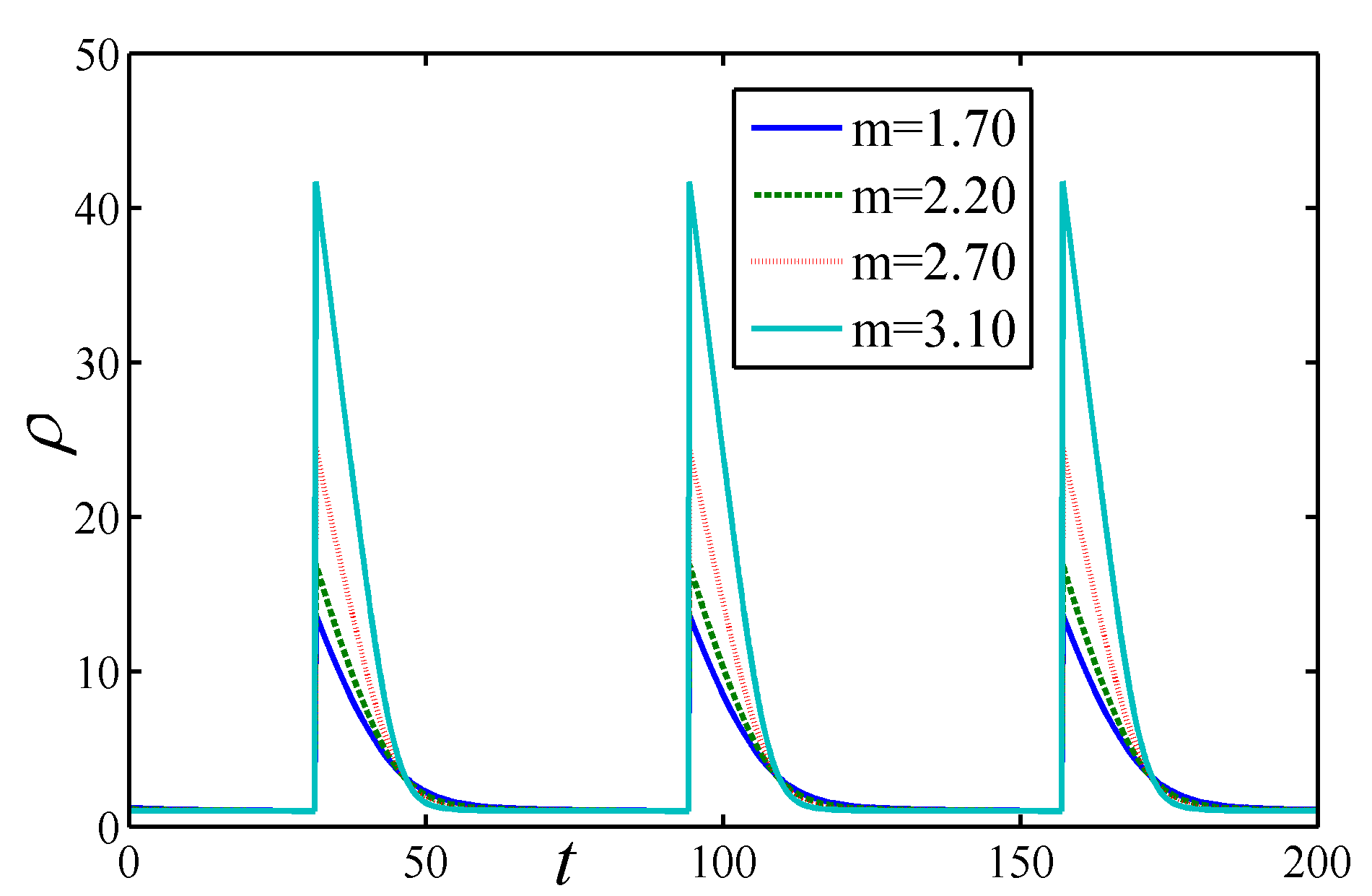}
					\caption{Energy density $\rho$ against cosmic time for $1.70\leq m \leq 3.20$ and $n=0.10$.  }
					\label{fig10}
				\end{minipage}
			\end{figure}

			Figure (\ref{fig9}) and Figure (\ref{fig10}) are representing the profile of pressure and energy density.  As per the evolution of time, energy density and pressure are positive and negative respectively in a periodic way for different values of m.
			\par
\begin{figure}[ht]
				\centering
				\begin{minipage}{.5\textwidth}
					\includegraphics[width=.8\linewidth]{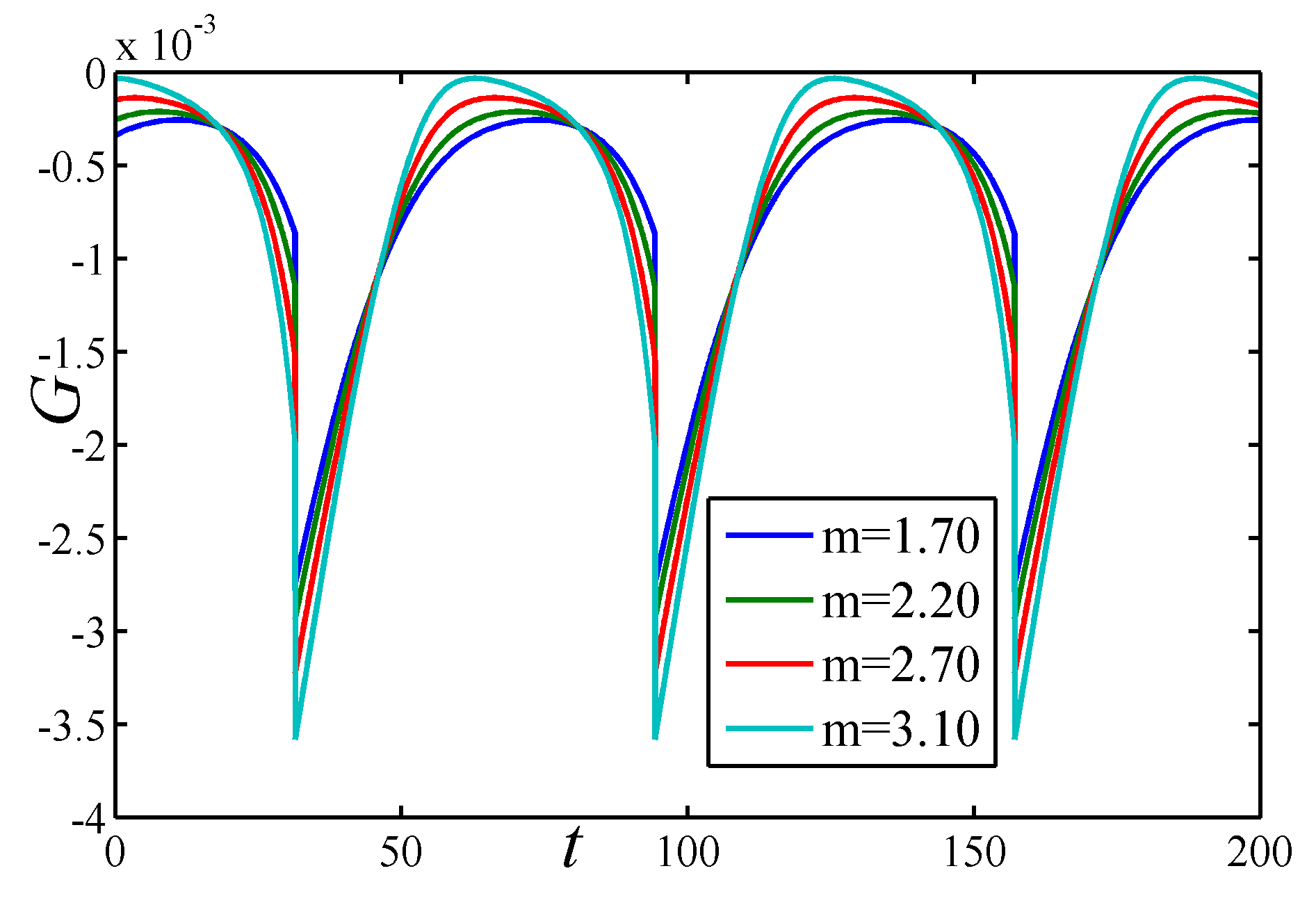}
					\caption{Gravitational constant $G$ against cosmic \newline time for $1.70\leq m \leq 3.20$,$c_7=c_8=A=1$, \newline $\alpha=0.5$,$c_4=55$ and $n=0.10$. }
					\label{fig11}
				\end{minipage}%
				\begin{minipage}{.5\textwidth}
					\includegraphics[width=.8\linewidth]{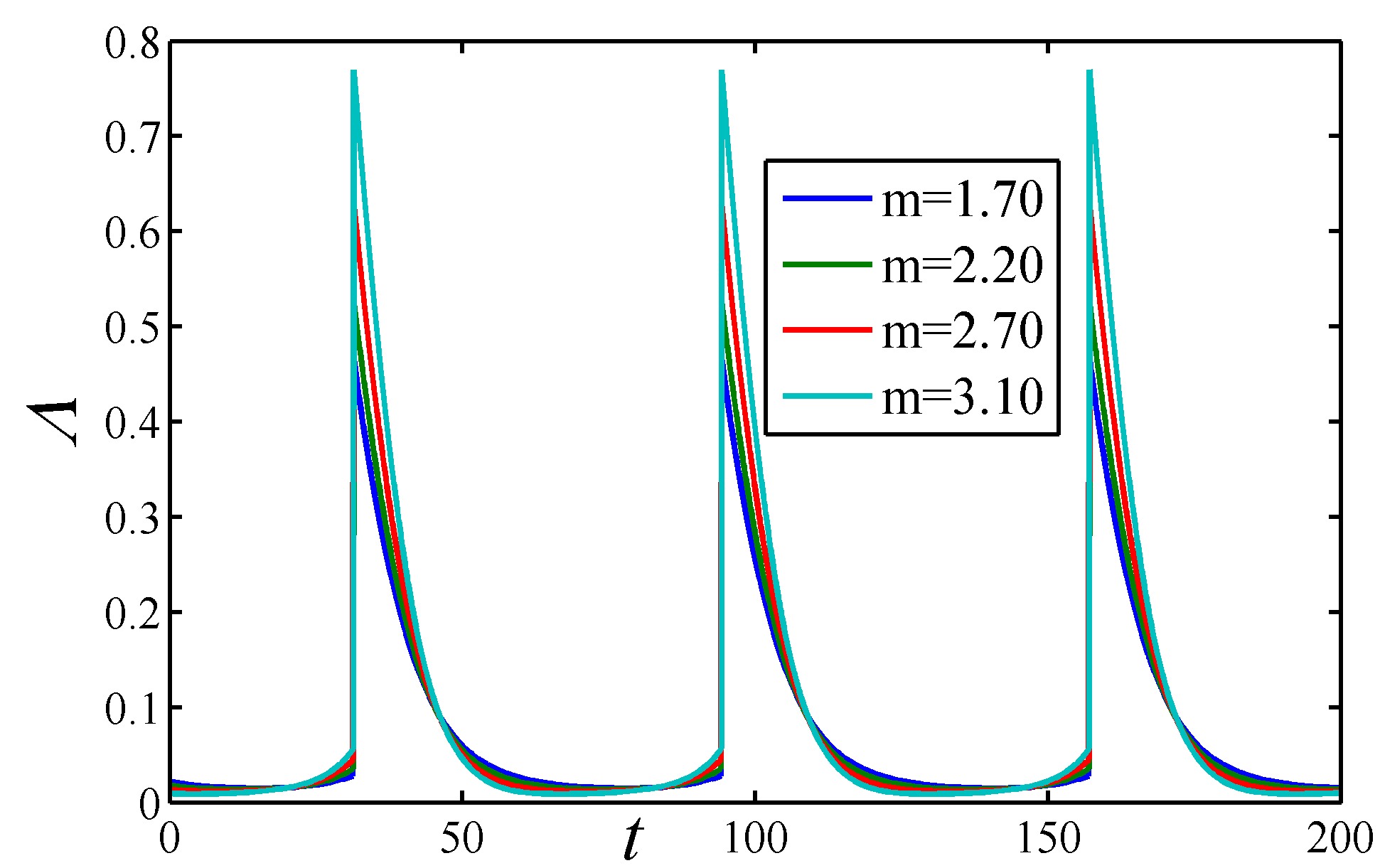}
					\caption{Cosmological constant $\Lambda$ against cosmic time for $1.70\leq m \leq 3.20$,$c_7=c_8=A=1$,$\alpha=0.5$,$c_4=55$ and $n=0.10$.}
					\label{fig12}
				\end{minipage}
			\end{figure}

			Figure (\ref{fig11}) and Figure (\ref{fig12}) shows the qualitative behaviour of gravitational constant and cosmological constant with respect to cosmic time. We noticed that, both behaves in a periodic way with time and also $\Lambda>0$ and $G<0$ with the evolution of cosmic time. $\Lambda>0$ and $G<0$ are free from the initial singularity.
			\par
\begin{figure}[ht!]
				\centering
				\begin{minipage}{.5\textwidth}
					\includegraphics[width=.8\linewidth]{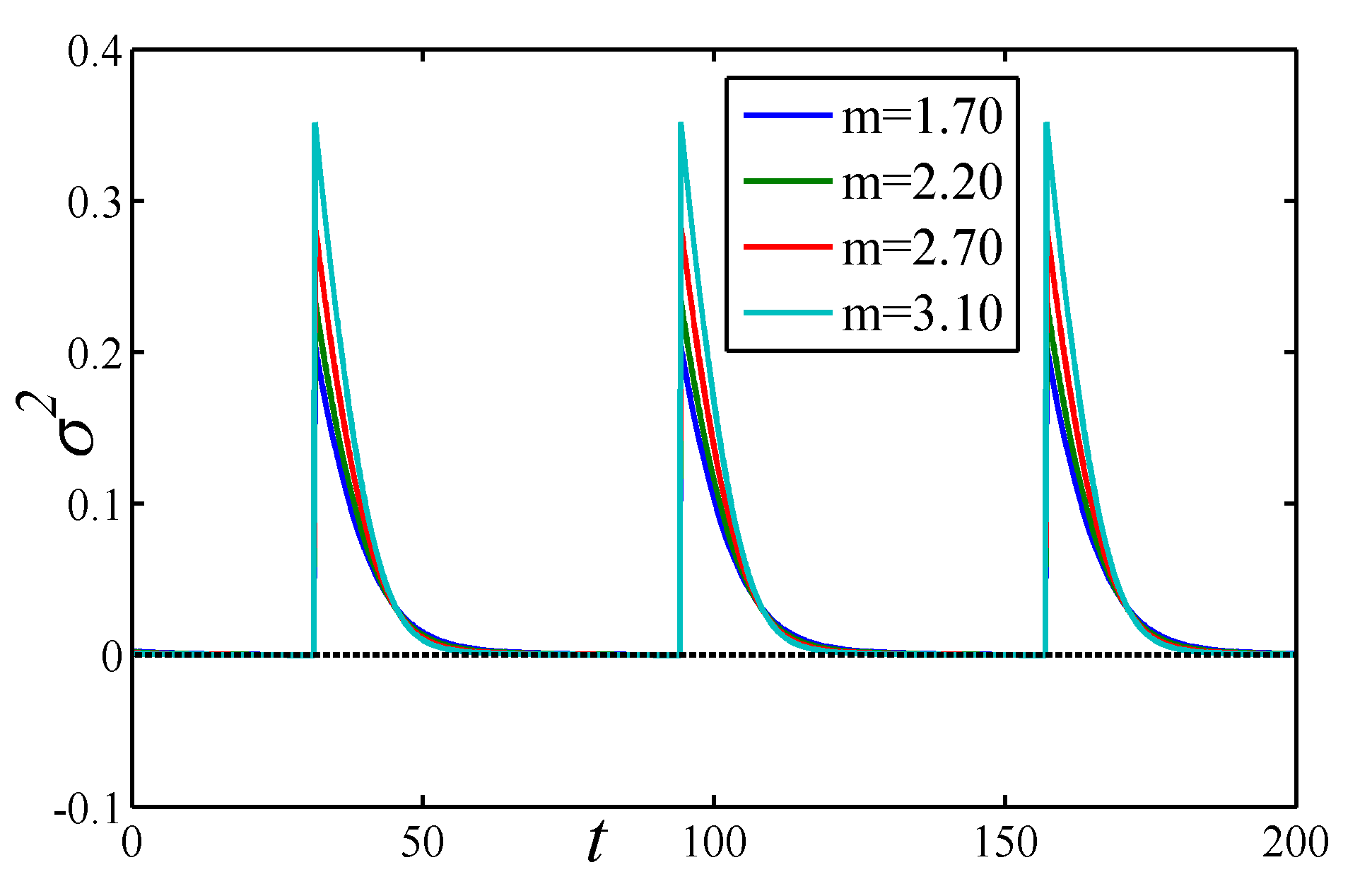}
					\caption{Shear scalar $\sigma^2$ against cosmic time \newline for $1.70\leq m \leq 3.20$,$c_7=c_8=A=1$,$\alpha=0.5$,\newline $c_4=55$ and $n=0.10$. }
					\label{fig13}
				\end{minipage}%
				\begin{minipage}{.5\textwidth}
					\includegraphics[width=.8\linewidth]{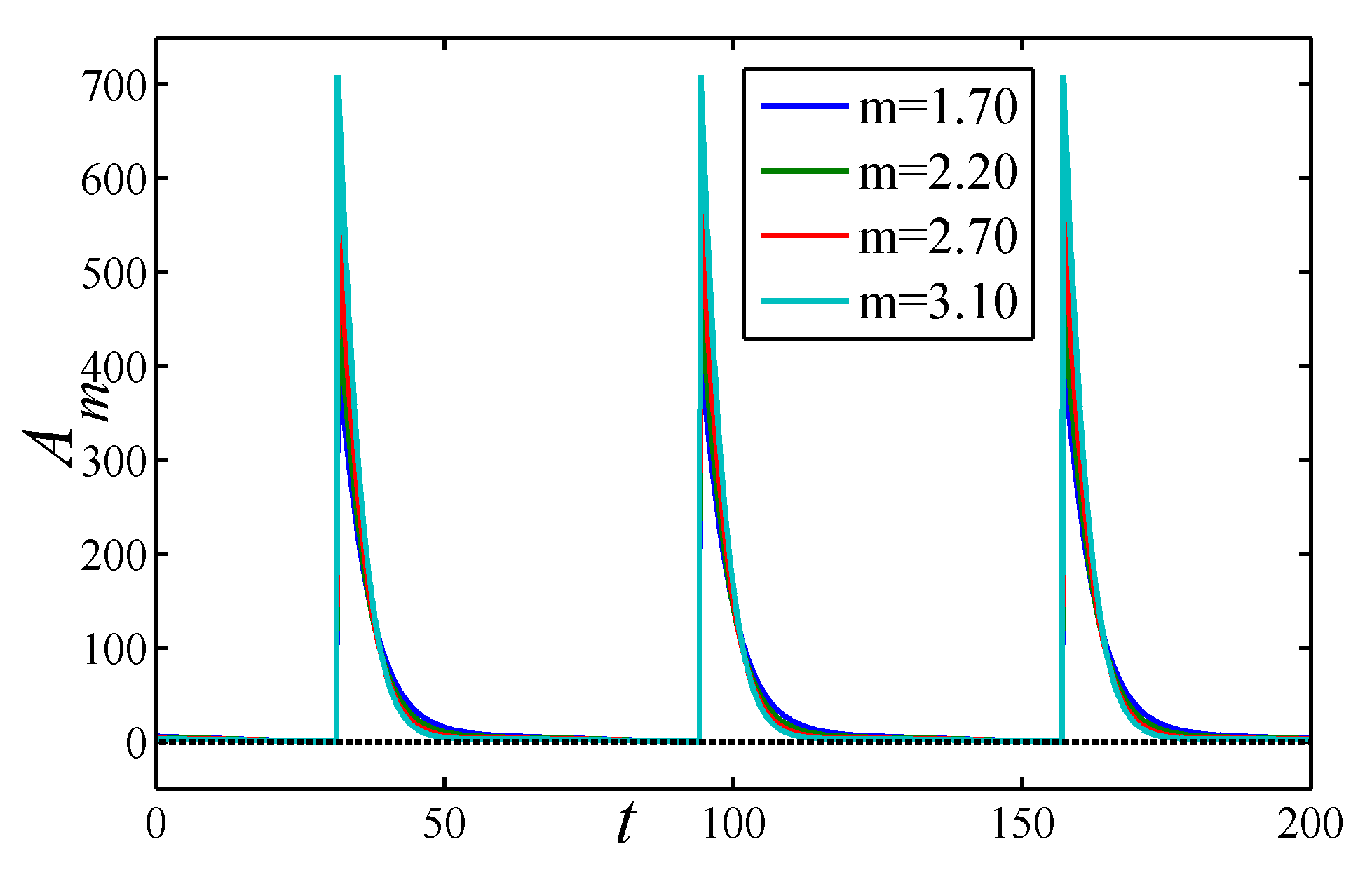}
					\caption{Anisotropic parameter $A_m$ against cosmic time for $1.70\leq m \leq 3.20$,$c_7=c_8=A=1$,$\alpha=0.5$,$c_4=55$ and $n=0.10$.}
					\label{fig14}
				\end{minipage}
			\end{figure}

			Figure (\ref{fig13}) and Figure (\ref{fig14}) shows the qualitative behaviour of shear scalar and anisotropic parameter with respect to cosmic time. Both the parameters behave in a periodic way with time and positive with the evolution of cosmic time.
			
			\begin{table}[ht]
				\centering
				\caption{$\{r,s\}$ pair for different model parameters with fixed $q$}
				{\begin{tabular}{@{}|c|c|c|c|c|c|@{}}
						\hline
						$q$& $n$ & $m$ & $c_4$ & $r$ & $s$ \\ \hline
						-0.5& 0.02 & 0.5358059184 & 250 & 1.000000001$\approx$ 1 & -3.33333333$\times 10^{-10}\approx 0$ \\ \hline
						-0.5& 0.04& 0.5097105689 & 250 & 0.9999999951$\approx$ 1& 1.633333333$\times 10^{-9}\approx 0$ \\ \hline
						-0.5& 0.06& 0.5043861337 & 250 & 1.000000008$\approx$ 1 & -2.666666667$\times 10^{-9}\approx 0$   \\ \hline
						-0.5& 0.08& 0.8416493515 & 10 & 1.000000001 $\approx$ 1& -3.333333333$\times 10^{-10}\approx 0$ \\
						\hline
					\end{tabular}}
					\label{Tab13}
				\end{table}
				
				\subsection{Case-III: $c_4<0 \;\text{i.e.\ } c_4=-c_5,\;c_5>0$}
\begin{figure}[ht]
					\centering
					\begin{minipage}{.5\textwidth}
						\centering
						\includegraphics[width=.8\linewidth]{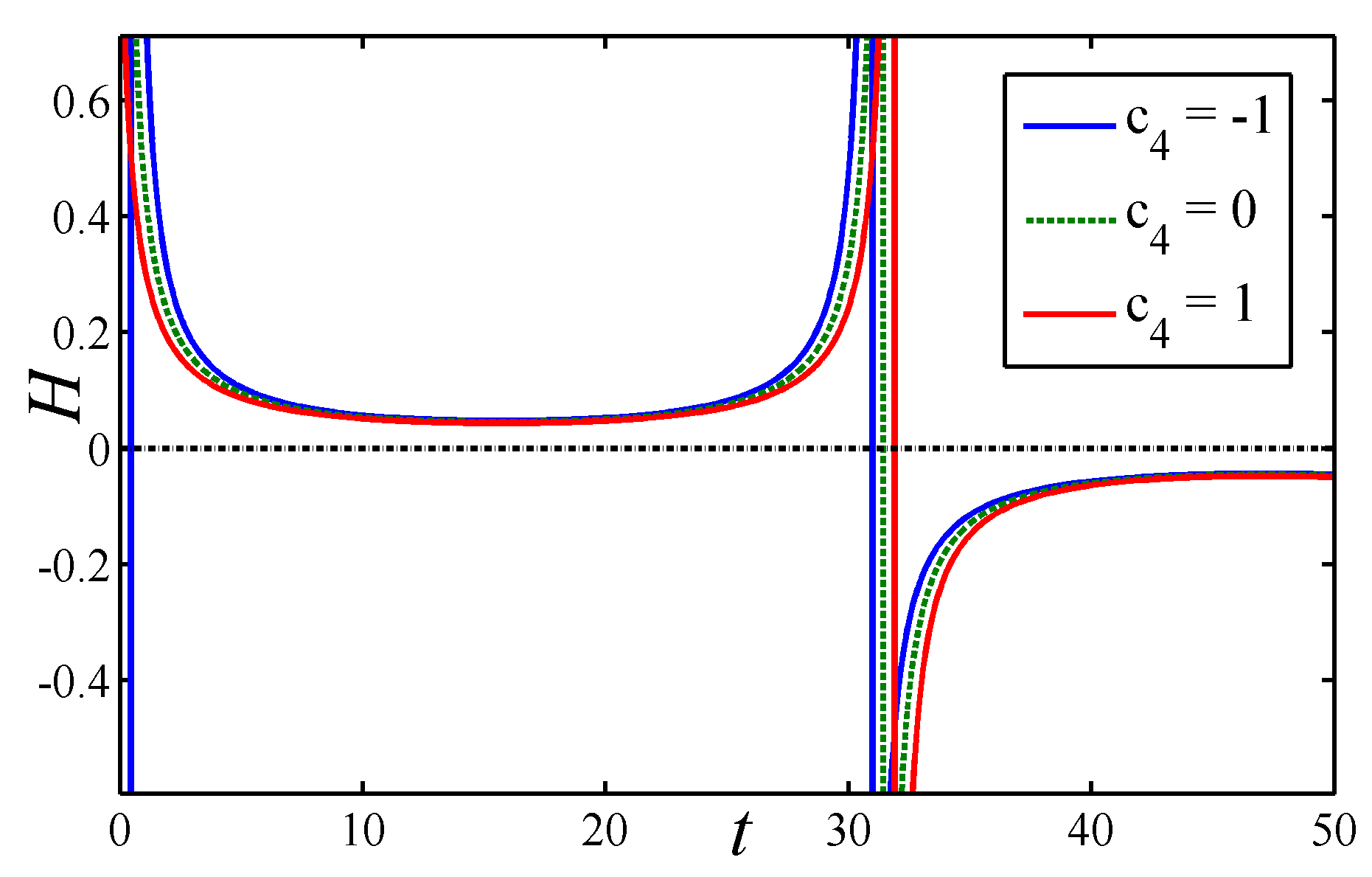}
						\caption{Hubble parameter $H$ against cosmic time \newline for $m=2.2$, $n=0.1$ and different $c_4$.}
						\label{fig15}
					\end{minipage}%
					\begin{minipage}{.5\textwidth}
						\centering
						\includegraphics[width=.8\linewidth]{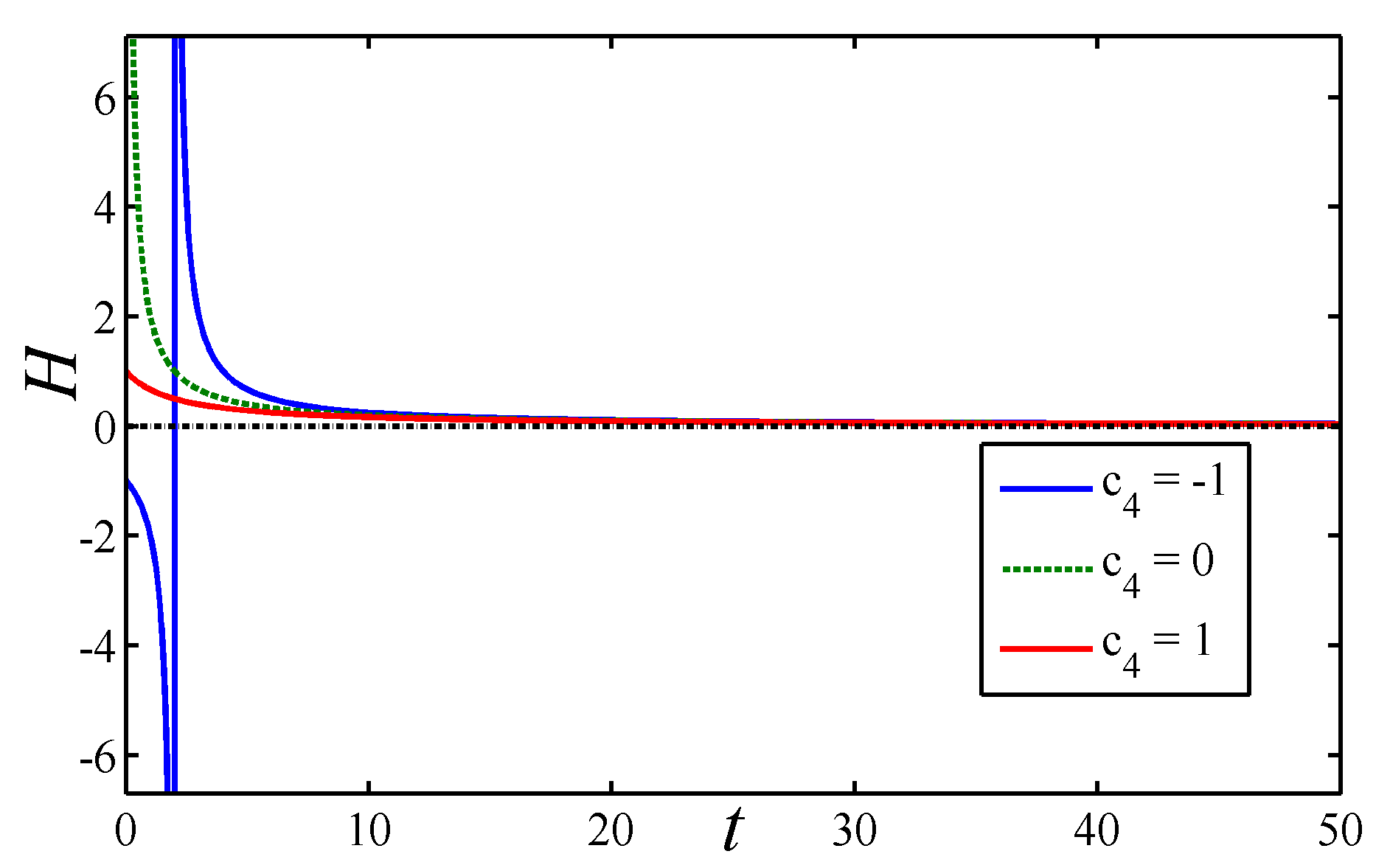}
						\caption{Hubble parameter $H$ against cosmic time for $m=0.5$, $n=0.01$ and different $c_4$.}
						\label{fig16}
					\end{minipage}
				\end{figure}

\begin{figure}[ht]
					\centering
					\begin{minipage}{.5\textwidth}
						\centering
						\includegraphics[width=.8\linewidth]{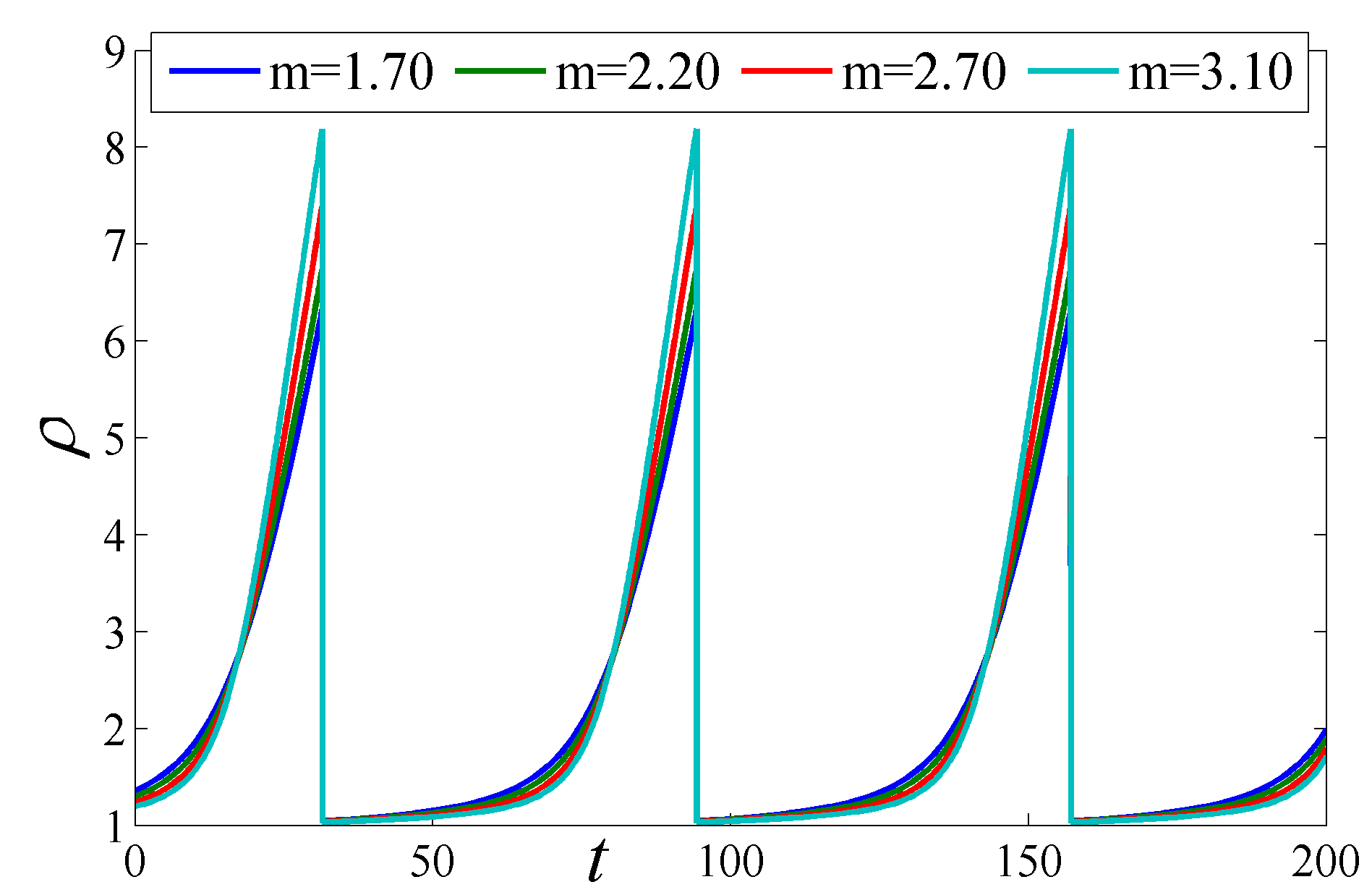}
						\caption{Energy density $\rho$ against cosmic time \newline for $1.70\leq m \leq 3.20$ and $n=0.10$.}
						\label{fig17}
					\end{minipage}%
					\begin{minipage}{.5\textwidth}
						\centering
						\includegraphics[width=.8\linewidth]{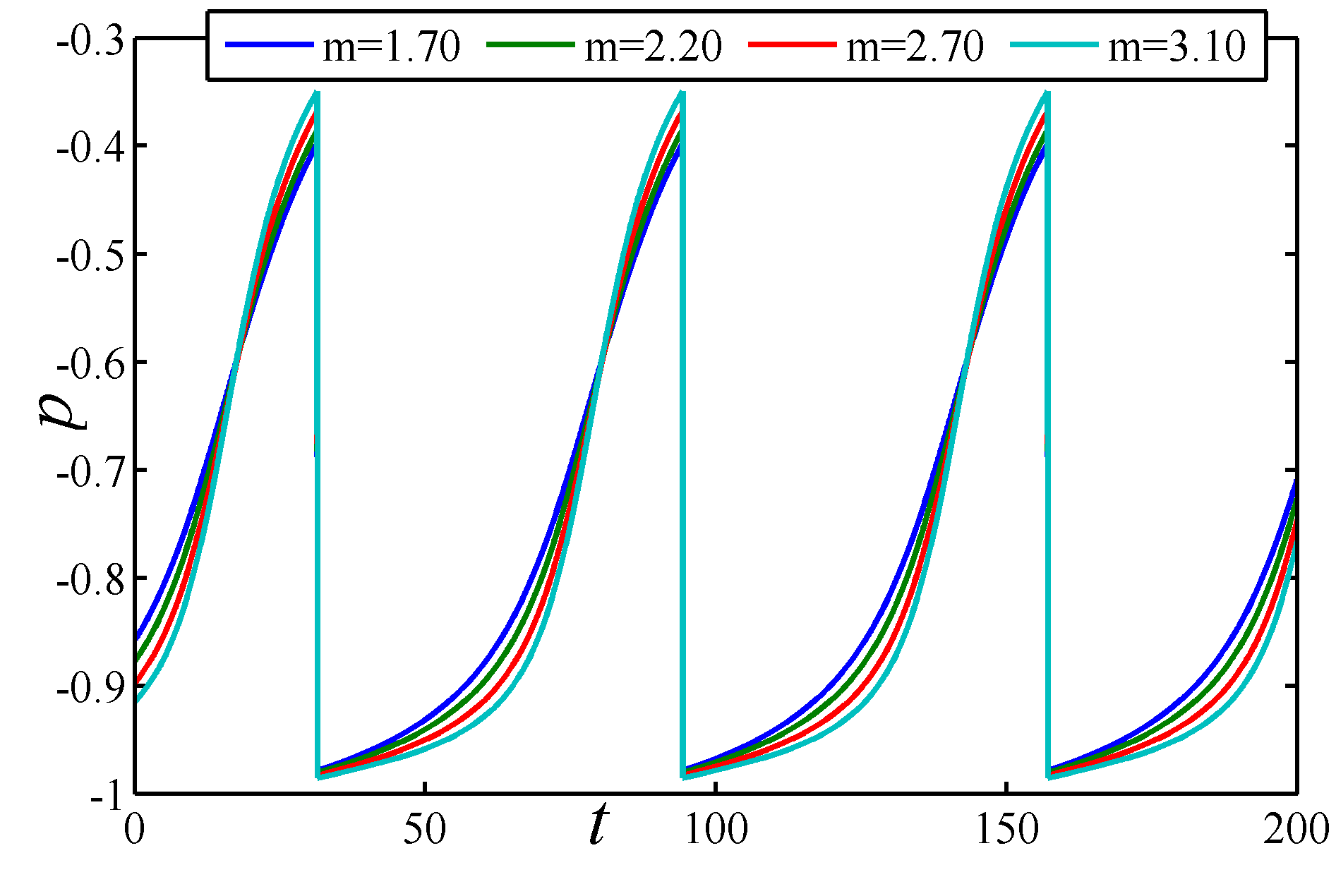}
						\caption{Pressure $p$ against cosmic time for $1.70\leq m \leq 3.20$ and $n=0.10$.  }
						\label{fig18}
					\end{minipage}
				\end{figure}

\begin{figure}[ht]
					\centering
					\begin{minipage}{.5\textwidth}
						\includegraphics[width=.8\linewidth]{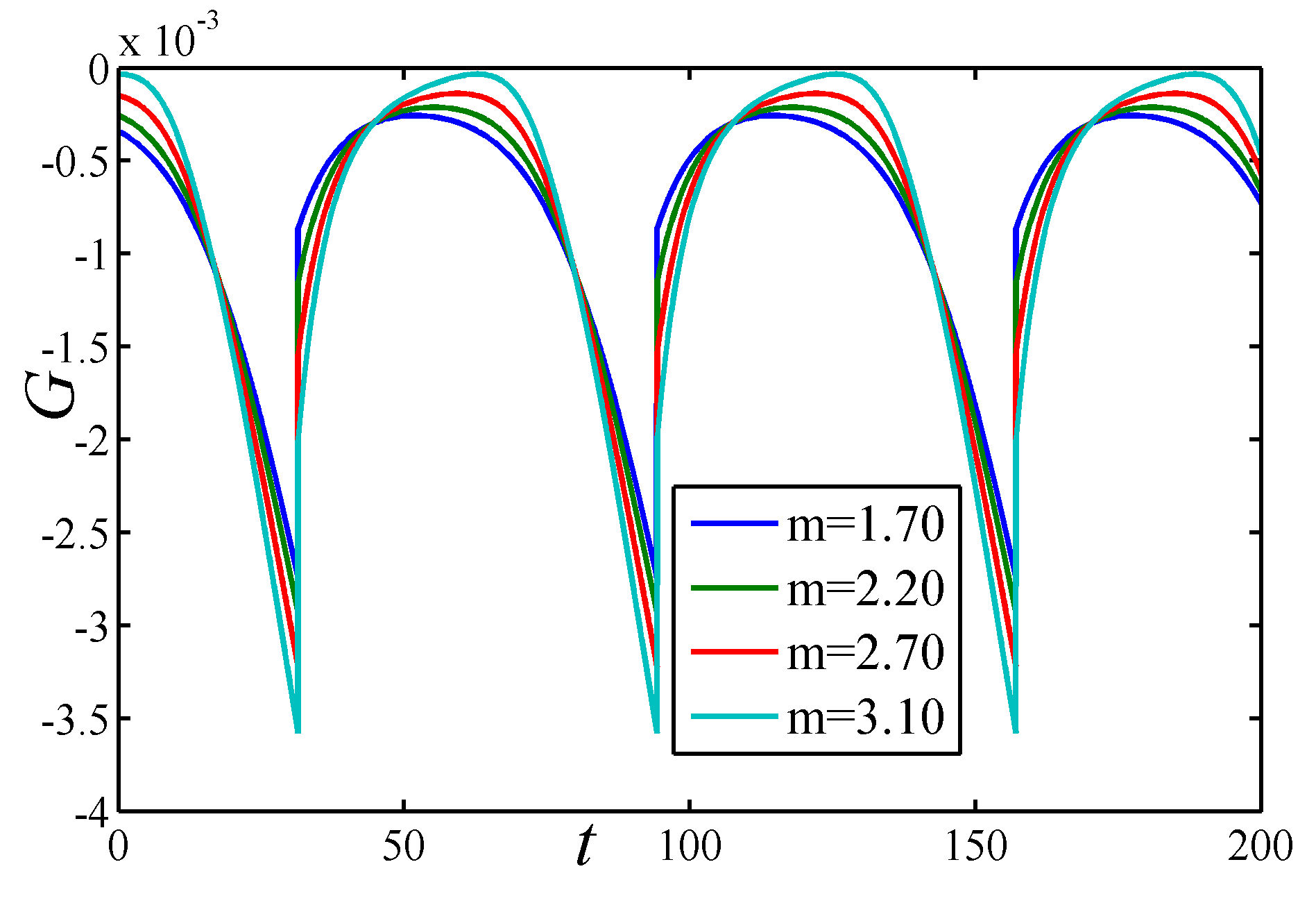}
						\caption{Profile of Gravitational constant $G$ \newline against cosmic time for $1.70\leq m \leq 3.20$,\newline $c_7=c_8=A=1$,$\alpha=0.5$,$c_4=55$ and $n=0.10$. }
						\label{fig19}
					\end{minipage}%
					\begin{minipage}{.5\textwidth}
						\includegraphics[width=.8\linewidth]{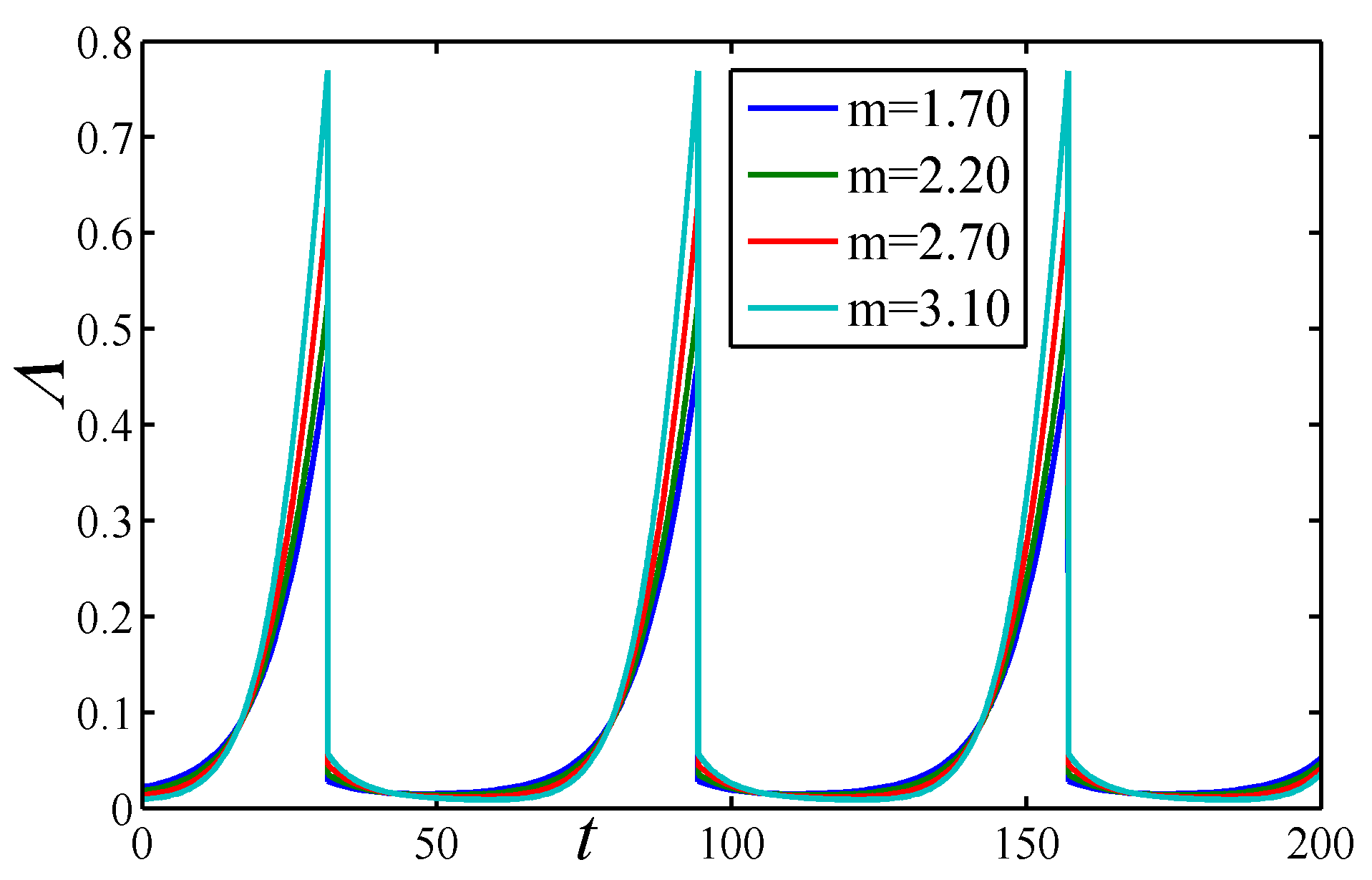}
						\caption{Cosmological constant $\Lambda$ against cosmic time for $1.70\leq m \leq 3.20$,$c_7=c_8=A=1$,$\alpha=0.5$,$c_4=55$ and $n=0.10$.}
						\label{fig20}
					\end{minipage}
				\end{figure}

\begin{figure}[ht]
					\centering
					\begin{minipage}{.5\textwidth}
						\includegraphics[width=.8\linewidth]{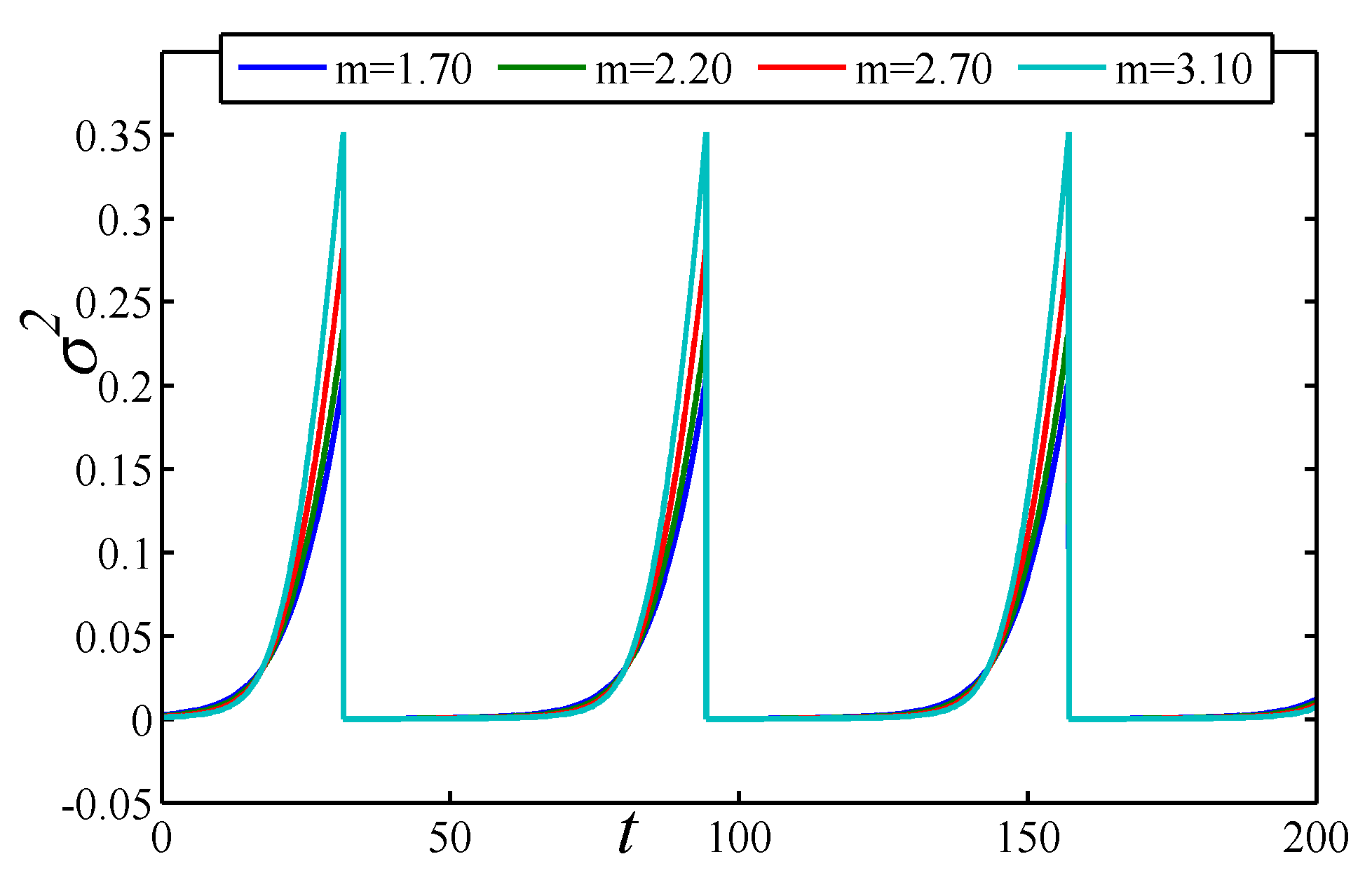}
						\caption{Shear scalar $\sigma^2$ against cosmic time \newline for $1.70\leq m \leq 3.20$,$c_7=c_8=A=1$,\newline $\alpha=0.5$,$c_4=55$ and $n=0.10$. }
						\label{fig21}
					\end{minipage}%
					\begin{minipage}{.5\textwidth}
						\includegraphics[width=.8\linewidth]{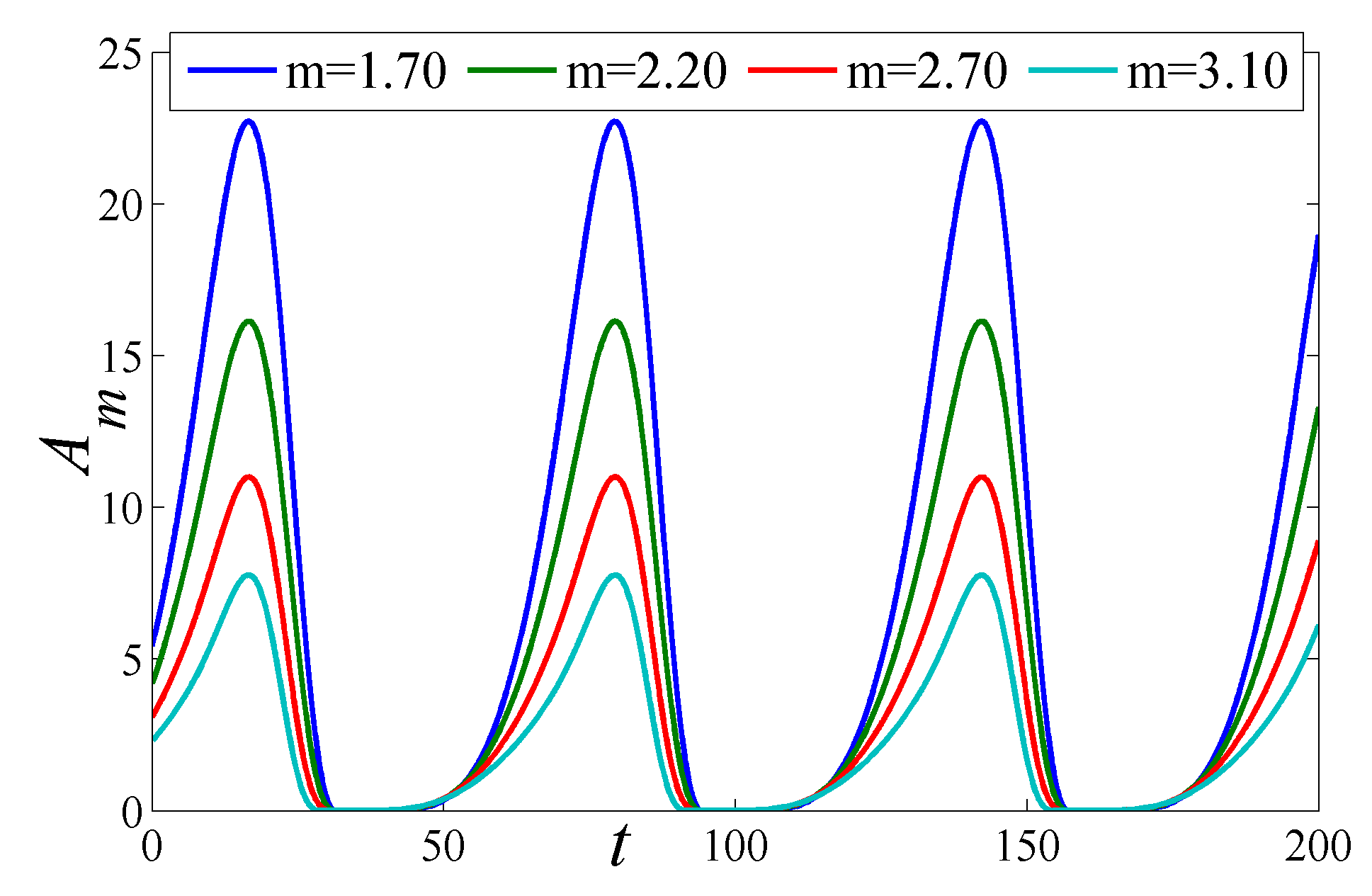}
						\caption{Anisotropic parameter $A_m$ against cosmic time for $1.70\leq m \leq 3.20$,$c_7=c_8=A=1$,$\alpha=0.5$,$c_4=55$ and $n=0.10$.}
						\label{fig22}
					\end{minipage}
				\end{figure}

				In this case the Hubble parameter in equation \eqref{e13} takes the form
				\begin{equation}\label{e47}
				H=\frac{n}{m\sin(nt)-nc_5}.
				\end{equation}
				The law $H=\frac{\dot{\mathcal{R}}}{\mathcal{R}}$ along with equation \eqref{e47} leads to the scale factor of the form
				\begin{equation}\label{e48}
				\mathcal{R}=c_9e^{-\frac{2}{\sqrt{n^2c_5^2-m^2}}\arctan\left(\frac{nc_5\tan\left(\frac{nt}{2}\right)-m}{\sqrt{n^2c_5^2-m^2}}\right)},\:\:c_9\; \text{is the constant of integration}.
				\end{equation}
				Further, equations \eqref{e11} and \eqref{e48} yields the following metric potentials as
				\begin{equation}\label{e49}
				R_i=c_{i5} e^{-\frac{2\arctan\left(\frac{nc_5\tan\left(\frac{nt}{2}\right)-m}{\sqrt{n^2c_5^2-m^2}}\right)}{\sqrt{n^2c_5^2-m^2}}}
				exp\left[c_{i6}\int e^{\frac{6\arctan\left(\frac{nc_5\tan\left(\frac{nt}{2}\right)-m}{\sqrt{n^2c_5^2-m^2}}\right)}{\sqrt{n^2c_5^2-m^2}}}dt\right],
				\end{equation}
				where $c_{i5}=c_ic_9$ and $c_{i6}=\frac{k_i}{3c_9^3} \: (i=1,2,3)$. The directional Hubble parameters are expressed as
				\begin{equation}\label{e50}
				H_i=\frac{\left[3nc_9^3e^{-\frac{6\arctan\left(\frac{nc_5\tan\left(\frac{nt}{2}\right)-m}{\sqrt{n^2c_5^2-m^2}}\right)}{\sqrt{n^2c_5^2-m^2}}}-nc_5k_i+mk_i\sin(nt)\right]e^{\frac{6\arctan\left(\frac{nc_5\tan\left(\frac{nt}{2}\right)-m}{\sqrt{n^2c_4^2-m^2}}\right)}{\sqrt{n^2c_5^2-m^2}}}}{3c_9^3(m\sin(nt)-nc_5)}.
				\end{equation}
				The expression for the energy density can be obtained from equations \eqref{e9} and \eqref{e21} as
				\begin{equation}\label{e51}
				\rho=\left[A+\frac{c_7}{\mathcal{R}^{3(1+\alpha)}}\right]^{\frac{1}{1+\alpha}}=\left[A+\rho_2e^{\frac{6(1+\alpha)\arctan\left(\frac{nc_5\tan\left(\frac{nt}{2}\right)-m}{\sqrt{n^2c_5^2-m^2}}\right)}{\sqrt{n^2c_5^2-m^2}}}\right]^{\frac{1}{1+\alpha}},
				\end{equation}
				where $\rho_2=c_7c_9^{-3(1+\alpha)}$. The expressions for gravitational and cosmological constants are obtained as
				\begin{equation}\label{e52}
				G=\frac{\splitfrac{-9n^2mc_9^6\left(\cos^2\left(\frac{nt}{2}\right)-\frac{1}{2}\right)e^{\frac{12\arctan\left(\frac{m-nc_5\tan\left(\frac{nt}{2}\right)}{\sqrt{n^2c_5^2-m^2}}\right)}{\sqrt{n^2c_5^2-m^2}}}+\bigg(m^2\cos^4\left(\frac{nt}{2}\right)}{-m^2\cos^2\left(\frac{nt}{2}\right)+nmc_5\sin\left(\frac{nt}{2}\right)\cos\left(\frac{nt}{2}\right)-\frac{n^2c_5^2}{4}}\bigg)k_7}{72\pi c_9^6(\rho+p)\left(\splitfrac{m^2\cos^4\left(\frac{nt}{2}\right)-m^2\cos^2\left(\frac{nt}{2}\right)}{+nmc_5\sin\left(\frac{nt}{2}\right)\cos\left(\frac{nt}{2}\right)-\frac{n^2c_5^2}{4}}\right)e^{\frac{12\arctan\left(\frac{m-nc_5\tan\left(\frac{nt}{2}\right)}{\sqrt{n^2c_5^2-m^2}}\right)}{\sqrt{n^2c_5^2-m^2}}}},
				\end{equation}
				where $k_7=\frac{k_4}{m}$,
				\begin{equation}\label{e53}
				\Lambda=\frac{\splitfrac{9n^2c_9^6\left(\rho m\left(\cos^2\left(\frac{nt}{2}\right)-\frac{1}{2}\right)-\frac{3}{4}(\rho+p)\right)e^{\frac{12\arctan\left(\frac{m-nc_5\tan\left(\frac{nt}{2}\right)}{\sqrt{n^2c_5^2-m^2}}\right)}{\sqrt{n^2c_5^2-m^2}}}}{+\bigg(m^2\cos^4\left(\frac{nt}{2}\right)-m^2\cos^2\left(\frac{nt}{2}\right)+nmc_5\sin\left(\frac{nt}{2}\right)\cos\left(\frac{nt}{2}\right)-\frac{n^2c_5^2}{4}\bigg)(k_5\rho+k_6 p)}}{9c_9^6(\rho+p)\left(\splitfrac{m^2\cos^4\left(\frac{nt}{2}\right)-m^2\cos^2\left(\frac{nt}{2}\right)}{+nmc_5\sin\left(\frac{nt}{2}\right)\cos\left(\frac{nt}{2}\right)-\frac{n^2c_5^2}{4}}\right)e^{\frac{12\arctan\left(\frac{m-nc_5\tan\left(\frac{nt}{2}\right)}{\sqrt{n^2c_5^2-m^2}}\right)}{\sqrt{n^2c_5^2-m^2}}}}.
				\end{equation}
				In this case, the physical quantities are obtained as
				\begin{equation}\label{e54}
				\Theta=\frac{3n}{m\sin(nt)-nc_5},
				\end{equation}
				\begin{equation}\label{e55}
				\sigma^2=\frac{(k_1^2+k_2^2+k_3^2)}{18c_9^6}e^{-\frac{12\arctan\left(\frac{m-nc_5\tan\left(\frac{nt}{2}\right)}{\sqrt{n^2c_5^2-m^2}}\right)}{\sqrt{n^2c_5^2-m^2}}},
				\end{equation}
				\begin{dmath}\label{e56}
					A_m=\frac{\splitfrac{27n^2m^2c_9^6\left(-2\tan \left(\frac{nt}{2}\right)+\tan^2 \left(\frac{nt}{2}\right)\sin(nt)+\sin(nt)\right)^2e^{-\frac{12\arctan\left(\frac{nc_5\tan\left(\frac{nt}{2}\right)-m}{\sqrt{n^2c_5^2-m^2}}\right)}{\sqrt{n^2c_5^2-m^2}}}}{+4\left(-\frac{nc_5}{2}-\frac{nc_5\tan\left(\frac{nt}{2}\right)}{2}+m\tan\left(\frac{nt}{2}\right)\right)^2(k_1^2+k_2^2+k_3^2)(m\sin(nt)-nc_5)^2e^{\frac{12\arctan\left(\frac{nc_5\tan\left(\frac{nt}{2}\right)-m}{\sqrt{n^2c_5^2-m^2}}\right)}{\sqrt{n^2c_5^2-m^2}}}}}{27n^2c_9^6\left(-nc_5-nc_5\tan^2\left(\frac{nt}{2}\right)+2m\tan\left(\frac{nt}{2}\right)\right)^2}.
				\end{dmath}
				In this case the state finder parameters are defined and expressed as
				\begin{equation}\label{e57}
				r=\frac{\dddot{\mathcal{R}}}{\mathcal{R}H^3}=\frac{(m\sin(nt)-nc_5)^3\left(\splitfrac{-4m^2\cos^4\left(\frac{nt}{2}\right)+(6m+4m^2)\cos^2\left(\frac{nt}{2}\right)}{+nmc_5\sin(nt)-3m-2m^2-1}\right)}{4m^2(m\sin(nt)-3nc_5)\cos^2\left(\frac{nt}{2}\right)\left(\cos^2\left(\frac{nt}{2}\right)-1\right)+n^3c_5^3-3mn^2c_5^2\sin(nt)},
				\end{equation}
				\begin{eqnarray}\label{e58}
					s&=&\frac{r-1}{3(q-0.5)}=\frac{2}{3(2m\cos(nt)-3)}\nonumber\\
					&\times&\left[\frac{(m\sin(nt)-nc_5)^3\left(\splitfrac{-4m^2\cos^4\left(\frac{nt}{2}\right)+(6m+4m^2)\cos^2\left(\frac{nt}{2}\right)}{+nmc_5\sin(nt)-3m-2m^2-1}\right)}{4m^2(m\sin(nt)-3nc_5)\cos^2\left(\frac{nt}{2}\right)\left(\cos^2\left(\frac{nt}{2}\right)-1\right)+n^3c_5^3-3mn^2c_5^2\sin(nt)}-1\right].
				\end{eqnarray}
				In terms of the deceleration parameter the state finder parameters are given as
				\begin{equation}\label{e59}
				r=\frac{\left(\sqrt{m^2-1-2q-q^2}-nc_5\right)^3\left(m^2-1-q+q^2-nc_5\sqrt{m^2-1-2q-q^2}\right)}{\left(m^2-1-2q-q^2+3n^2c_5^2\right)\sqrt{m^2-1-2q-q^2}-3nc_5\left(m^2-1-2q-q^2+\frac{n^2c_5^2}{3}\right)},
				\end{equation}
				\begin{eqnarray}\label{e60}
					s&=&\frac{2}{3(2q-1)}\times\nonumber\\
					&&\left[\frac{\left(\sqrt{m^2-1-2q-q^2}-nc_5\right)^3\left(m^2-1-q+q^2-nc_5\sqrt{m^2-1-2q-q^2}\right)}{\left(m^2-1-2q-q^2+3n^2c_5^2\right)\sqrt{m^2-1-2q-q^2}-3nc_5\left(m^2-1-2q-q^2+\frac{n^2c_5^2}{3}\right)}
					-1\right].\nonumber\\
				\end{eqnarray}
				Figure (\ref{fig15}) and Figure (\ref{fig16}) indicates the profile of Hubble parameter against cosmic time for different values of $c_4$. Similar qualitative behaviour as that of case-II is observed for physical parameters likes energy density[Figure (\ref{fig17})], pressure[Figure (\ref{fig18})], gravitational constant [Figure (\ref{fig19})], cosmological constant [Figure (\ref{fig20})], shear scalar [Figure (\ref{fig21})] and anisotropic parameter [Figure (\ref{fig22})].
				\par
				Further, we have analyzed whether these derived models are approaching $\Lambda$CDM model or not for the computational range of $n$ and $m$ (Table \ref{Tab11}). From equations \eqref{e31},\eqref{e32},\eqref{e45}, \eqref{e46}, \eqref{e59} and \eqref{e60}, we have evaluated the $\{r,s\}$ pair for the different model parameters, which are presented in  Table \ref{Tab12} to Table \ref{Tab14} respectively for three cases. Here one can notice that, model discussed in case-I approaches to $\Lambda$CDM model for different value of $q$ and $m$.  At current value of $q=-\frac{1}{2}$, the model in case-I approaches to $\Lambda$CDM model for $m=\frac{\sqrt{5}}{2}$. The models in case-II and Case-III, the $\{r,s\}$ pair depends on the model parameters $n$, $m$, $c_4$, $c_5$ and $q$. At current value of $q=-\frac{1}{2}$, the model in case-II approaches to $\Lambda$CDM model whereas model in Case-III does not approaches to $\Lambda$CDM model in view of the values of $n$ and $m$ presented in Table \ref{Tab11}.
				
				\begin{table}[ht]
					\centering
					\caption{$\{r,s\}$ pair for different model parameters with fixed $q$}
					{\begin{tabular}{@{}|c|c|c|c|c|c|@{}}
							\hline
							$q$& $n$ & $m$ & $c_5$ & $r$ & $s$ \\ \hline
							-0.5& 0.02 & 1.127031913 & 1 & 1.000000001$\approx 1$  & -3.333333333$\times 10^{-10}\approx 0$ \\ \hline
							-0.5& 0.04& 1.136137315 & 1 & 1.020403999& -0.006801333000\\ \hline
							-0.5& 0.06& 1.144350162 & 1 & 1.041217996 & -0.01373933200   \\ \hline
							-0.5& 0.08& 1.154670505 & 1 & 1.062447982& -0.01373933200 \\
							\hline
						\end{tabular}}
						\label{Tab14}
					\end{table}
					\section{Stability and physical acceptability of the solutions}
					\subsection{The squared sound speed}
					We determine the classical stability of considered models on the basis of an adiabatic squared sound speed. It is one of the important quantity in cosmology. Adiabatic squared sound speed for system is defined \cite{28}, as
					\begin{equation}\label{e61}
					c_s^2=\frac{\partial p}{\partial \rho}.
					\end{equation}
					Here $c_s^2$ has three possibilities i.e $c_s^2<0$ or $c_s^2=0$ or $c_s^2>0$. The sign of $c_s^2$ is very important to investigate as it leads to the instability of the cosmological models through which one can reject or accept the constructed cosmological models. The case when, $c_s^2<0$, leads to classical instability of the cosmological models due to the uncontrolled grow of the energy density perturbation. The case when $c_s^2>0$ may leads to the issue of occurrence of  casuality. As a matter of fact, it is usually considered as $c_s\leq 1$ and the bound on $c_s^2$ is $0\leq c_s^2\leq 1$. In addition to that, the complementary bound $c_s>1$ is used as a condition for rejecting the theories. The details regarding $c_s^2$ can be found in \cite{22}.
					\par
\begin{figure}[ht!]
						\minipage{0.32\textwidth}
						{\includegraphics[width=\linewidth]{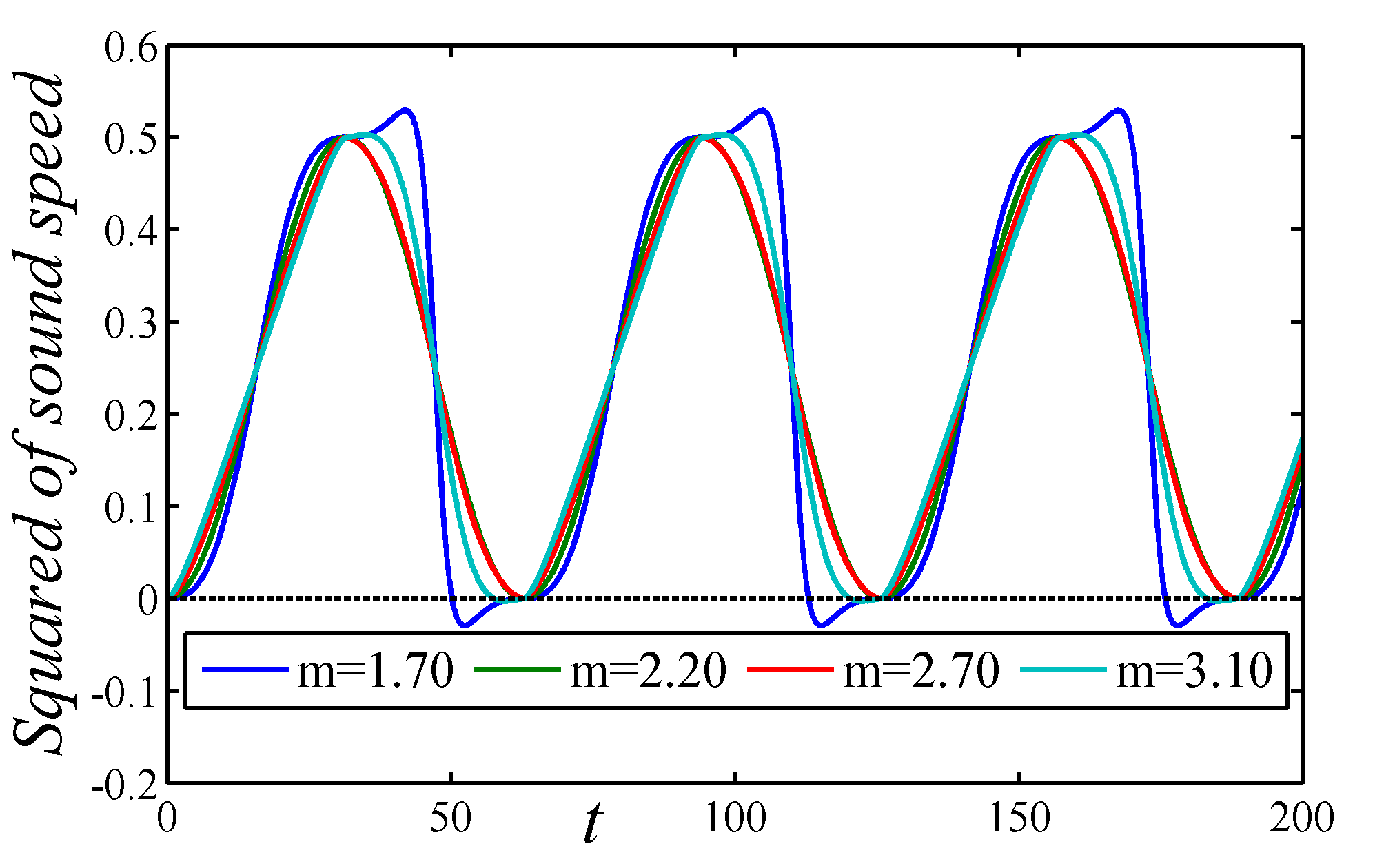}}
						\endminipage\hfill
						\minipage{0.32\textwidth}
						{\includegraphics[width=\linewidth]{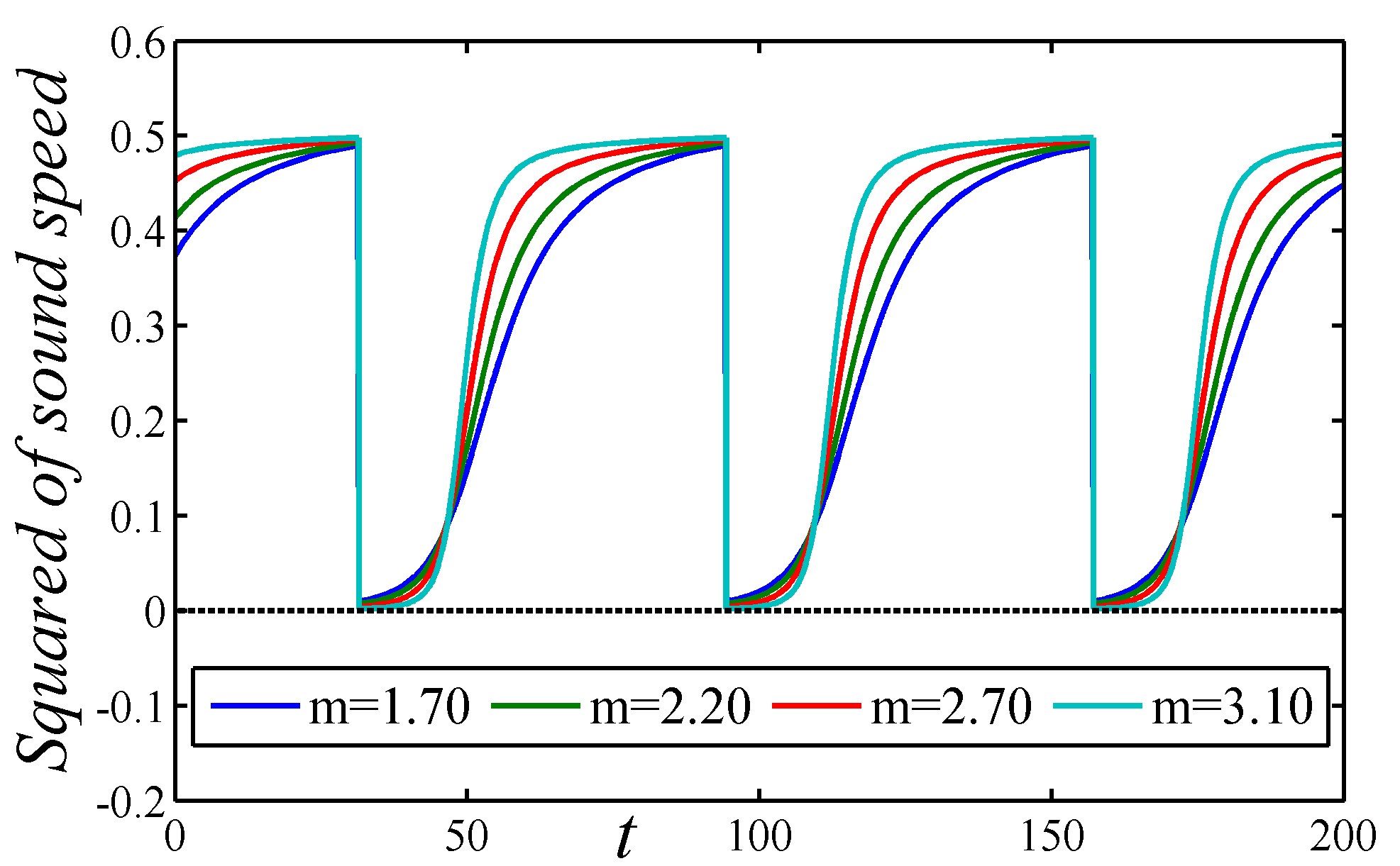}}
						\endminipage\hfill
						\minipage{0.32\textwidth}%
						{\includegraphics[width=\linewidth]{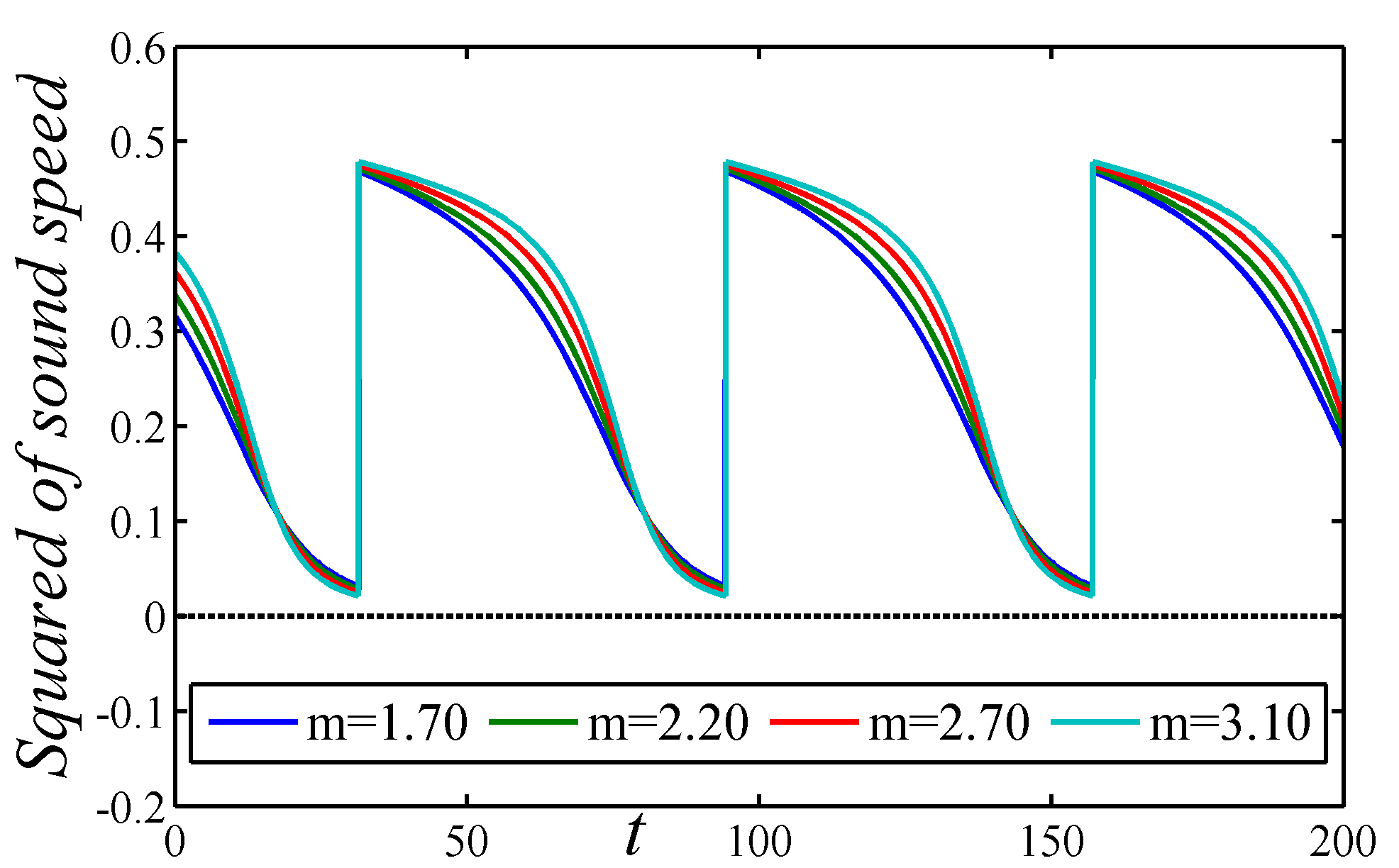}}
						\endminipage
						\caption{Squared sound speed against cosmic time for three different cases: $c_4=0$ (Left),$c_4>0$ (Right) and $c_4<0$ (Middle).}
						\label{fig23}
					\end{figure}

					In present study, $c_s^2$ is obtained as
					\begin{equation}\label{e62}
					c_s^2=\frac{\partial p}{\partial \rho}=\frac{A\alpha}{\rho^{1+\alpha}}=
					\begin{cases}
					\frac{A\alpha}{A+\rho_0\left(\sin(nt)\right)^{-\frac{3(1+\alpha)}{m}}\left(1+\cos(nt)\right)^{\frac{3(1+\alpha)}{m}}},& \text{for } c_4=0 \\[20pt]
					\frac{A\alpha}{A+\rho_1e^{-\frac{6(1+\alpha)\arctan\left(\frac{nc_4\tan\left(\frac{nt}{2}\right)+m}{\sqrt{n^2c_4^2-m^2}}\right)}{\sqrt{n^2c_4^2-m^2}}}}           ,& \text{for}\;c_4>0 \\[20pt]
					\frac{A\alpha}{A+\rho_2e^{\frac{6(1+\alpha)\arctan\left(\frac{nc_5\tan\left(\frac{nt}{2}\right)-m}{\sqrt{n^2c_5^2-m^2}}\right)}{\sqrt{n^2c_5^2-m^2}}}},&
					\text{for}\;c_4<0
					\end{cases}\ .
					\end{equation}
					In the Figure (\ref{fig23}), we have presented the profile of squared sound speed against time for different values of $m$. It is observed from the Figure (\ref{fig23}) that, $c_s^2$ is periodic in nature and $0\leq c_s^2<0.6$ in all the cases. Thus in account of the prescribe range of $c_s^2$, the solution presented here are stable.
					\subsection{Energy conditions}
\begin{figure}[ht!]
						\minipage{0.32\textwidth}
						{\includegraphics[width=\linewidth]{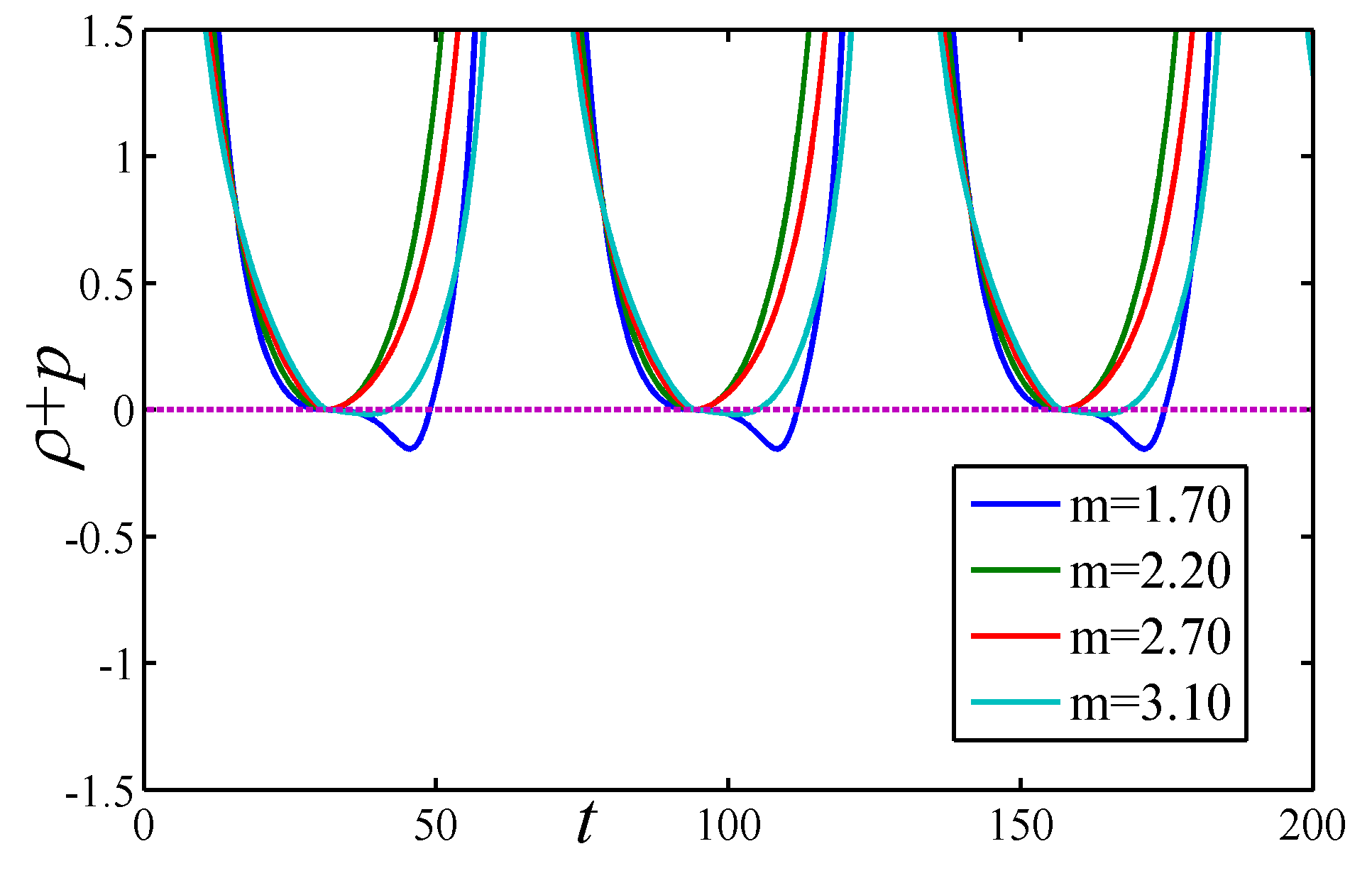}}
						\endminipage\hfill
						\minipage{0.32\textwidth}
						{\includegraphics[width=\linewidth]{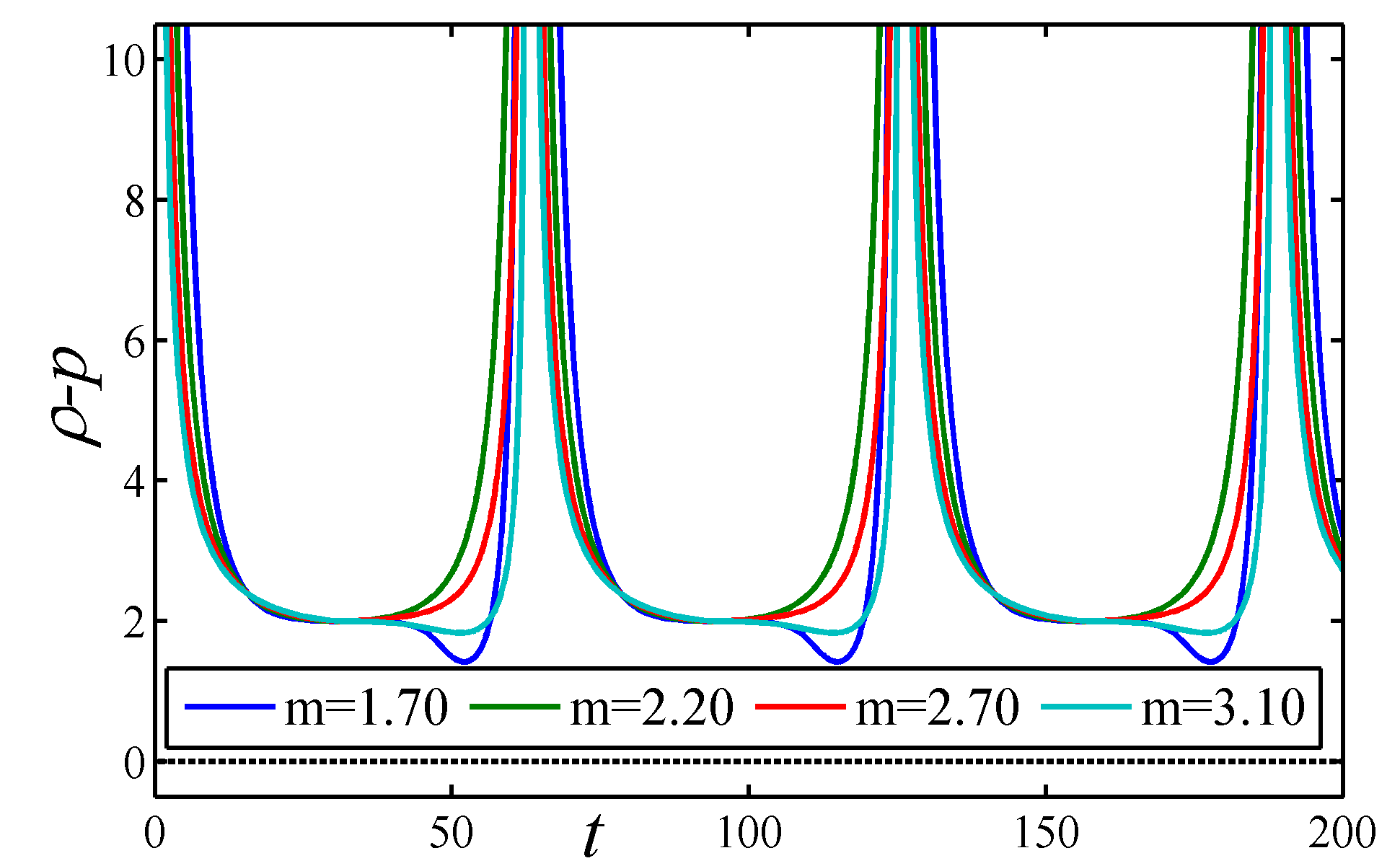}}
						\endminipage\hfill
						\minipage{0.32\textwidth}%
						{\includegraphics[width=\linewidth]{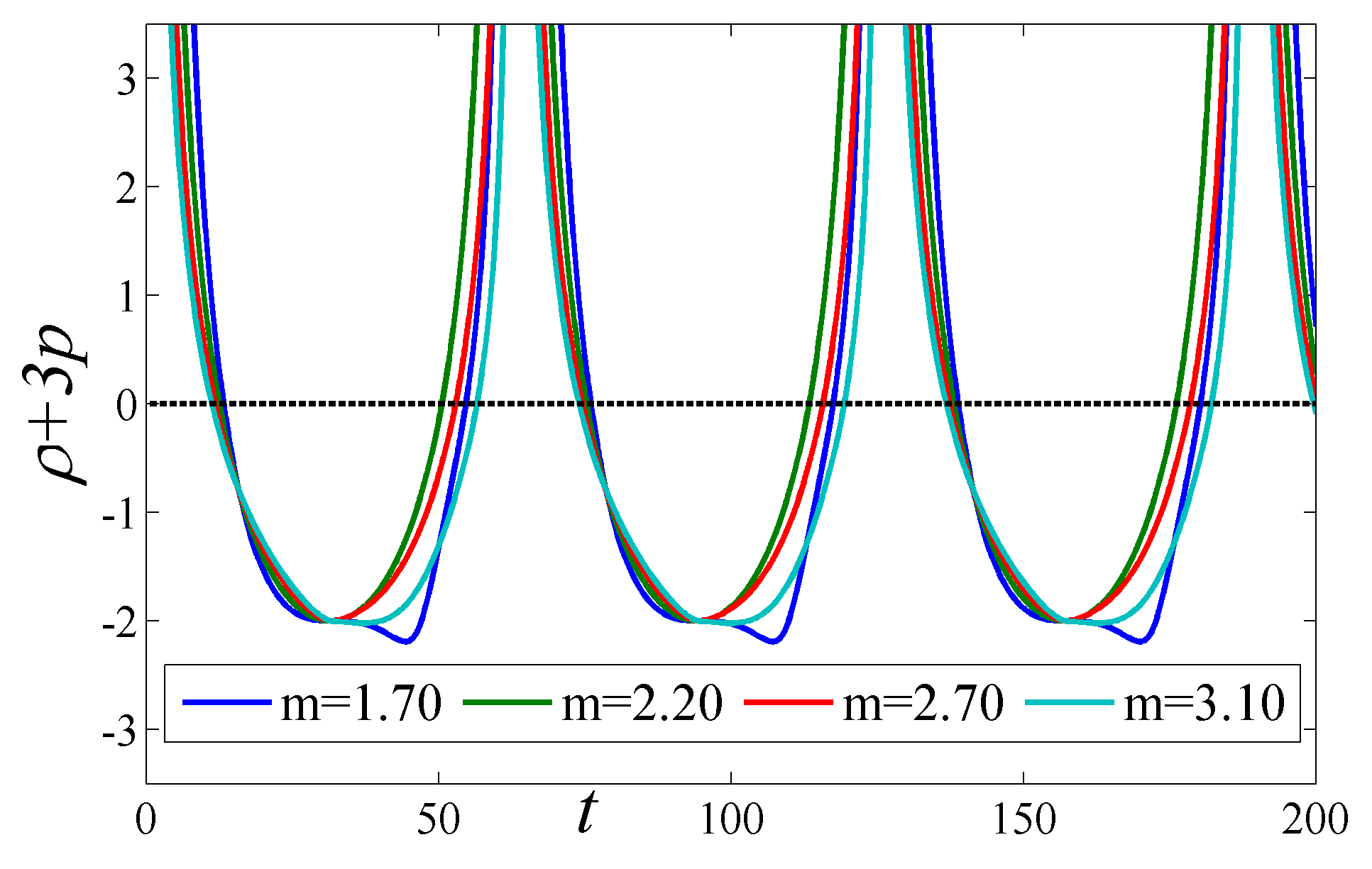}}
						\endminipage
						\caption{Energy conditions (WEC(Left), DEC (Middle), SEC (Right)) against time for case-I.}
						\label{fig24}
					\end{figure}

\begin{figure}[ht!]
						\minipage{0.32\textwidth}
						{\includegraphics[width=\linewidth]{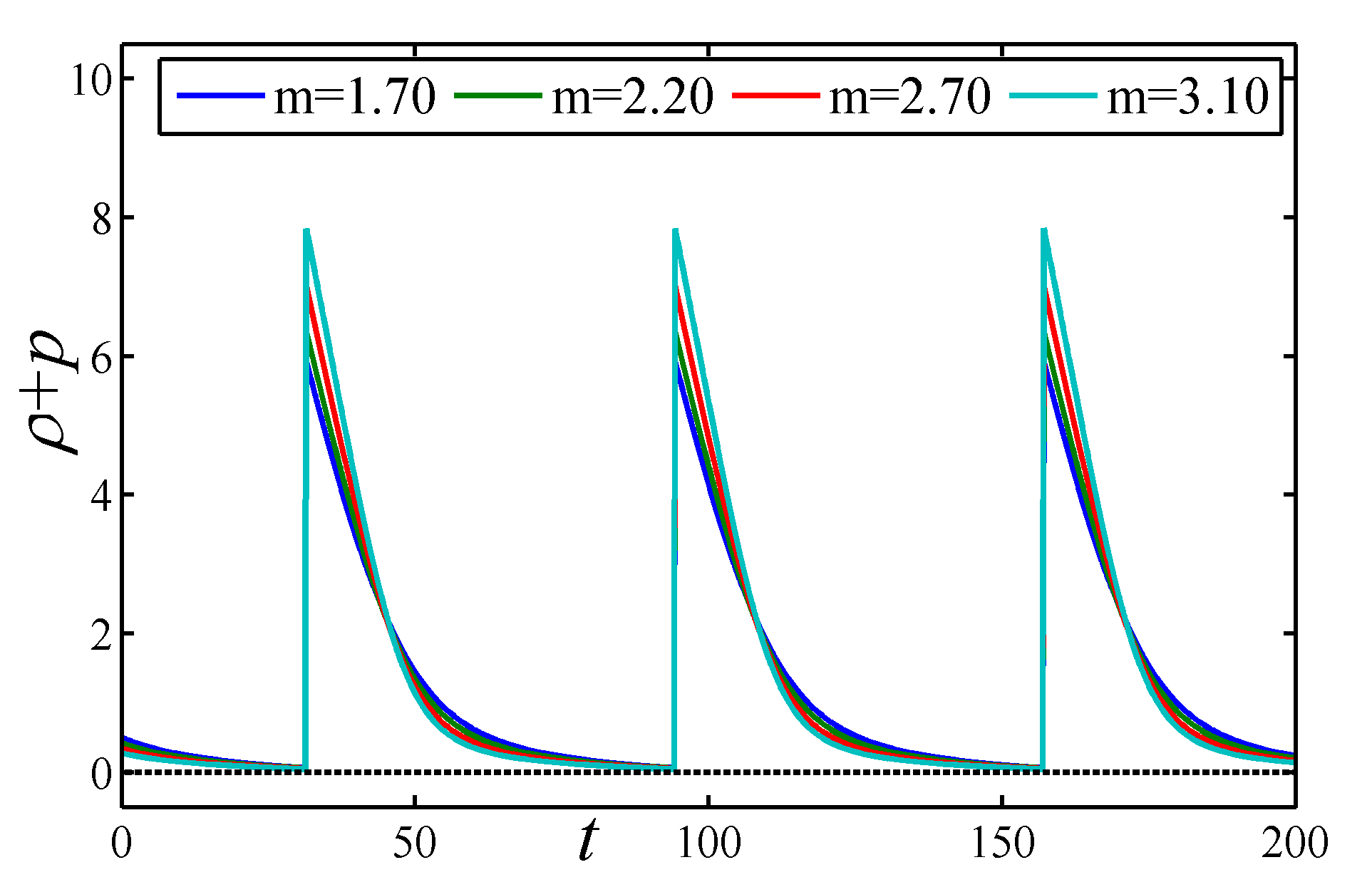}}
						\endminipage\hfill
						\minipage{0.32\textwidth}
						{\includegraphics[width=\linewidth]{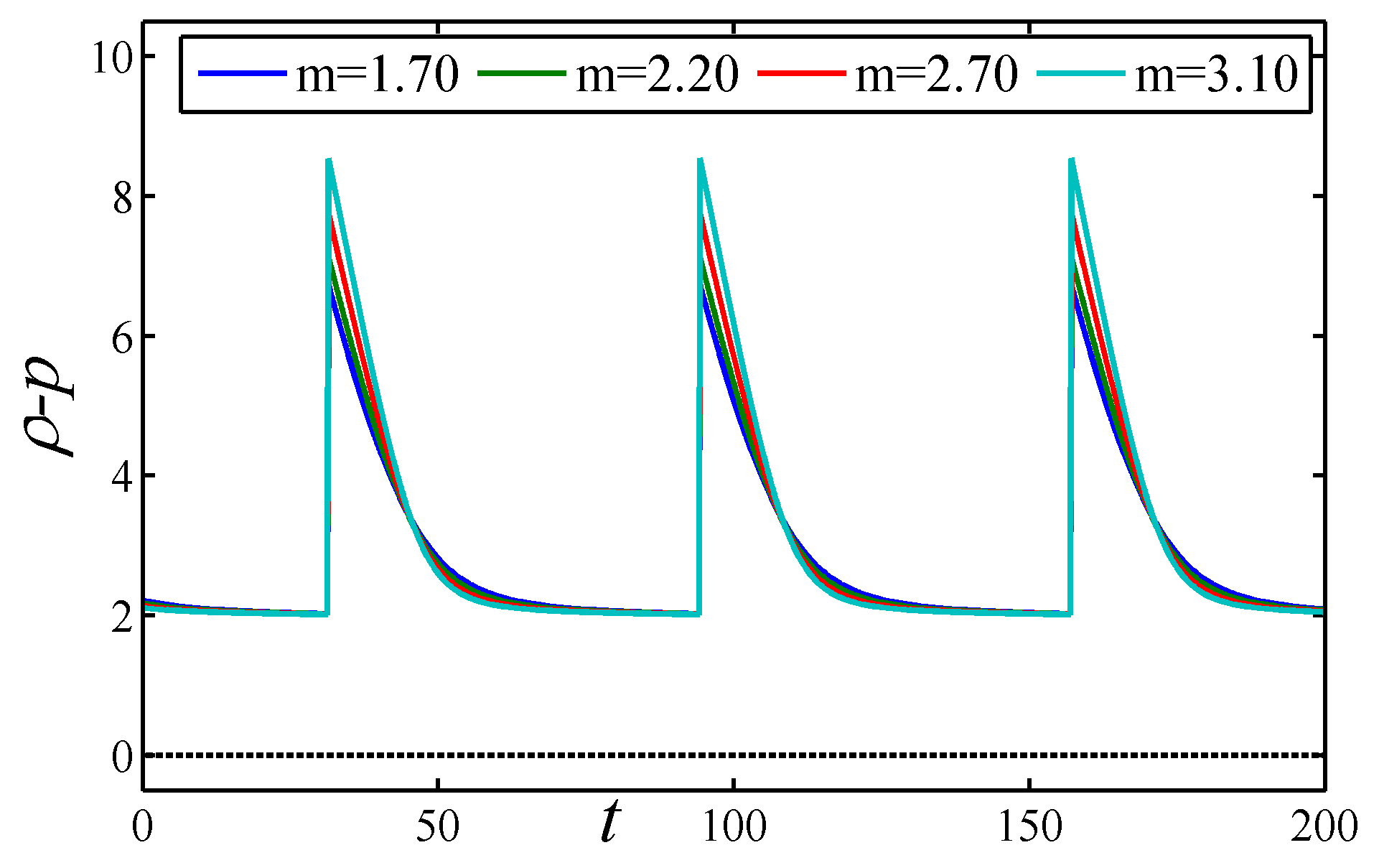}}
						\endminipage\hfill
						\minipage{0.32\textwidth}%
						{\includegraphics[width=\linewidth]{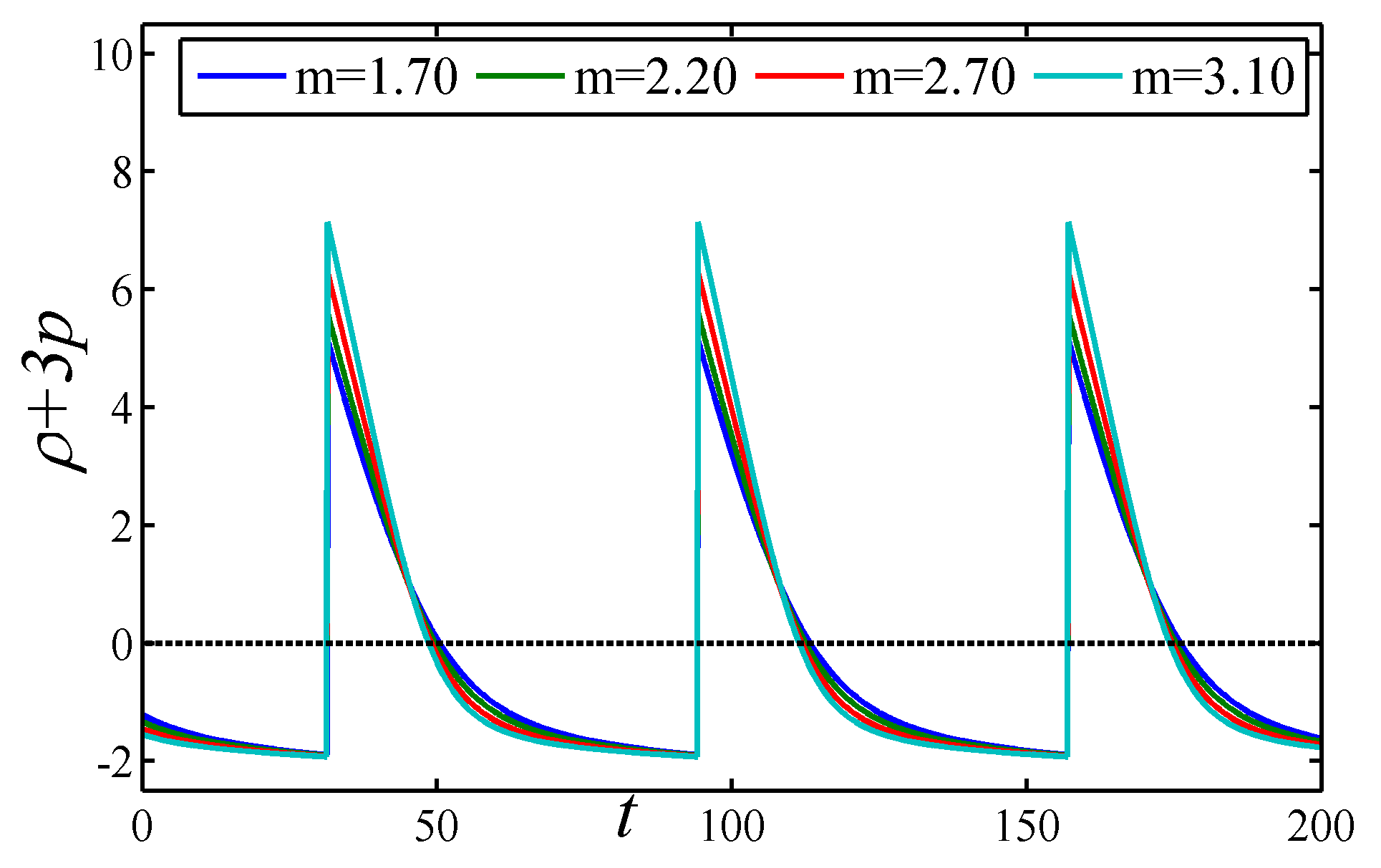}}
						\endminipage
						\caption{Energy conditions (WEC(Left), DEC (Middle), SEC (Right)) against time for case-II.}
						\label{fig25}
					\end{figure}

\begin{figure}[ht!]
						\minipage{0.32\textwidth}
						{\includegraphics[width=\linewidth]{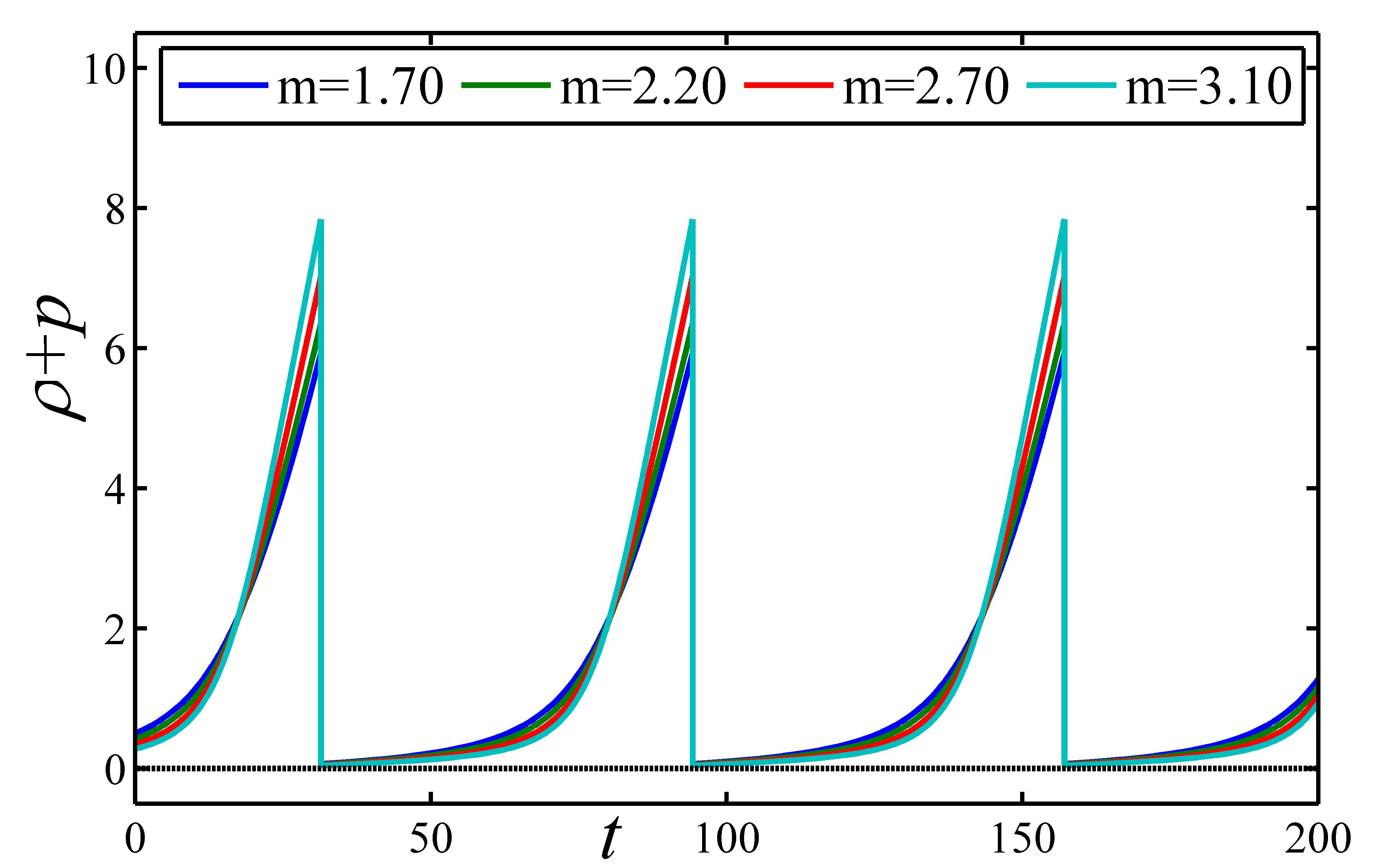}}
						\endminipage\hfill
						\minipage{0.32\textwidth}
						{\includegraphics[width=\linewidth]{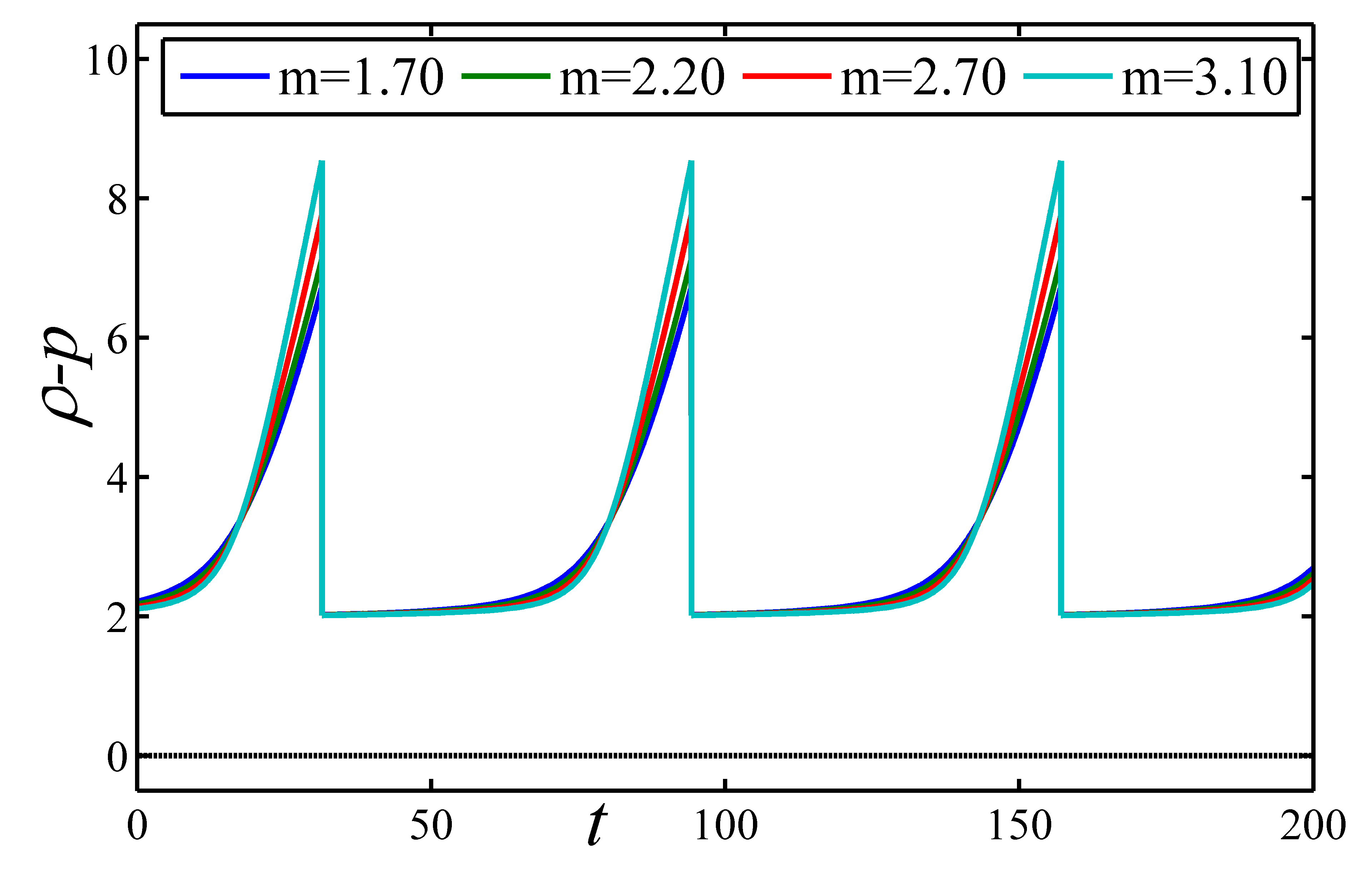}}
						\endminipage\hfill
						\minipage{0.32\textwidth}%
						{\includegraphics[width=\linewidth]{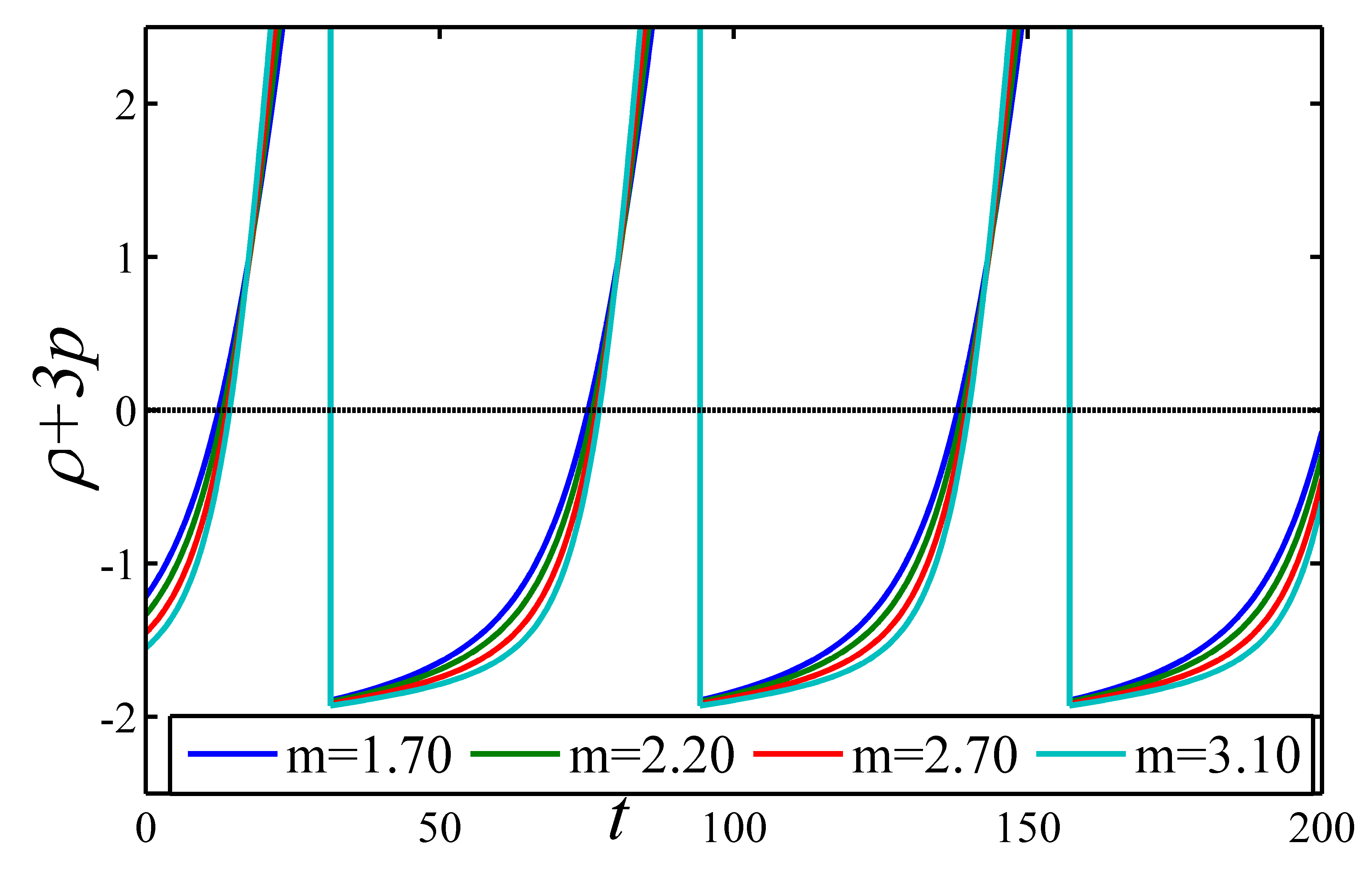}}
						\endminipage
						\caption{Energy conditions (WEC(Left), DEC (Middle), SEC (Right)) against time for case-III.}
						\label{fig26}
					\end{figure}

					The pointwise energy conditions (ECs) depends on energy momentum tensor at a given point in a space time. These conditions are contractions of time like or null vector with the Einstein's tensor and energy momentum tensor coming from Einstein's field equations. ECs can be imposed in order to investigate the constraints on the free parameters involved in the cosmological models. For example the evolution of acceleration or deceleration of the universe and the emergence of Big Rip singularity, can be related to the constraints imposed by the ECs. For energy momentum tensor $T_{ij}=(\rho +p)u_iu_j-pg_{ij}$, the standard pointwise energy conditions are defined as follows \cite{29,30,31}
					\begin{itemize}
						\item Null energy condition (NEC): $\rho+p \geq 0$
						\item Weak energy condition (WEC): $\rho \geq 0$ and $\rho+p \geq 0$
						\item Dominant energy condition (DEC): $\rho \geq 0$ and $\rho \pm p \geq 0$
						\item Strong energy condition (SEC): $\rho+p \geq 0$ and $\rho+3p \geq 0$
					\end{itemize}
					We use the ECs to investigate the stability and physical acceptibility of the solutions in the present models. In general the universe in cosmological model should satisfy WEC and DEC and violates SEC for late time accelerated expansion of the universe. For each of the three cases, Figures (\ref{fig24}), (\ref{fig25}) and (\ref{fig26}) represents the profile of WEC, DEC and SEC respectively versus cosmic time. From these figures it can seen that profile of energy conditions follows periodic variation for each of the cases. Considered model satisfies WEC $(\rho+p)$ (except for m=1.7 in $case-I$) and DEC $(\rho-p)$ whereas SEC is satisfied and violated periodically throughout the evolution of the universe.
					
\section{Conclusion}
In this work, we have considered Bianchi-I cosmological models with generalised Chaplygin gas  represented by equation (\ref{e17}) and cosmological solutions are obtained by using time periodic deceleration parameter represented by equation (\ref{e12}). By fixing the value of constant parameter $n$, we have computed the range of parameter $m$ (see Table-\ref{Tab11} for present observational value (range) of deceleration parameter). Figures (\ref{fig1}) and (\ref{fig2}) shows the variation of deceleration parameter with respect to cosmic time for $n=0.01$ and $n=0.10$ and corresponding range of $m$. For $n=0.01$ deceleration parameter is negative throughout the evolution of the universe whereas for $n=0.10$ it shows transition from decelerated phase to accelerated phase periodically throughout the evolution of the universe. Hubble parameter represented by (\ref{e13}) is computed by using it's relation with deceleration parameter. The integration constant $c_4$ affects the scale factor so we have discussed three different cases for the considered model depending on the positive negative and neutral nature of $c_4$. In each of the cases we have analyzed the solutions and  physical quantities of the observational interest (such as expansion scalar ($\Theta$), shear scalar ($\sigma^2$), anisotropy parameter ($A_m$), state finder parameters ($r$ and $s$)).
\par
Gravitational constant $G$ varies from positive to negative in the present cosmological framework. With positive and negative $G$, we will have attractive and repulsive nature of gravity respectively. Positive and negative cosmological constant also yields repulsive and attractive gravity respectively. The dark energy is responsible for accelerating universe and positive cosmological constant acts as a mechanism for dark energy. The periodic nature of $\Lambda$ and $G$ also highlight the attractive as well as repulsive gravity era in the model. For more details one can refer Ayuso et al.\cite{55}. In bouncing model, the scale factor is having its minima at the bounce instant. The universe should be contracting before the bounce and expanding after the bounce. The solutions given in Eq. \eqref{e34} and \eqref{e48} may lead to asymmetric bounce for different values of model parameters. On the other hand, the solution \eqref{e15} will not exhibit bouncing scenario, however, the toy model will exhibit periodic behaviour of scale factor but the behaviour is not of bouncing type.
\par
In the recent, Sahoo et al. \cite{25} have discussed the periodic varying deceleration parameter in $f(R,T)=R+2\lambda T$ gravity. For $\lambda=0$, this reduces to the result in general relativity. In our investigation, we noticed that for all the models SEC is violated but WEC and SEC are violated in their study. For stability of the solutions we both have different approaches, they have used linear homogeneous perturbations in the FRW background whereas we used the square sound speed. Our solutions are stable but their stability of solution depends on the parameter $k$ and $\lambda$. Further, all other physical parameters we both have the same periodic qualitative behaviour.
In all the figures, we have taken the cosmic time $t$ along the horizontal axis which is measured in giga years $(1 Gyr=10^9 years)$.  The conclusion of our study are as follows:
					\begin{itemize}
						
						\item Almost all the parameters which we have discussed in our study shows periodic behaviour due to the choice of deceleration parameter.
						\item For each of the cases energy $\rho$ is positive and $p$ is negative throughout the evolution of the universe and this negative pressure guarantees the late time expansion of the universe.
						\item Models discussed in Case-I and Case-II, tends to $\Lambda$CDM model whereas model in Case-III fails for the value of $n$ and $m$ provided in  Table \ref{Tab11}. The expressions obtained for expansion scalar, shear scalar, and anisotropic parameter bears singularities in Case-I whereas these are free from singularities in Case-II and Case-III.
						\item The square sound speed satisfies the bounds $0\leq c_s^2\leq 1$ for each of the cases. So the considered model is stable and physically acceptable.
						\item For each of the cases considered model satisfies WEC (except for m=1.7 in $case-I$) and DEC whereas it violate SEC periodically throughout the evolution of the universe. So using the results of energy conditions one can conclude that the violation of SEC may leads to the accelerating universe.
					\end{itemize}
\section*{Acknowledgement}
The authors are thankful to the reviewers for the constructive comments, which have certainly improved the quality of the manuscript. The author B. K. Bishi thankful to research supported wholly/in part by the National Research Foundation of South Africa (Grant Numbers: 118511) in the form of a postdoctoral fellowship at university of Zululand, South Africa.
					
				\end{document}